# Triboelectric Pixels as building blocks for microscale and large area integration of drop energy harvesters

Ali Ghaffarinejad,[+a] Xabier García-Casas,[+a]* Fernando Núñez-Gálvez,[a,b] Jorge Budagosky,[a,b] Vanda Godinho,[a] Carmen López-Santos,[a,b] Juan Ramón Sánchez-Valencia,[a] Ángel Barranco,[a] Ana Borrás[a]*

Nanotechnology on Surfaces and Plasma Laboratory, Consejo Superior de Investigaciones Científicas (CSIC), Materials Science Institute of Seville (CSIC-US). c/ Américo Vespucio 49, 41092, Seville (Spain)

Departamento de Física Aplicada I, Universidad de Sevilla, C/ Virgen de Africa 7, 41011, Seville (Spain)

+These two authors have contributed equally

anaisabel.borras@icmse.csic.es; xabier.garcia@icmse.csic.es

**SUMMARY**

The ultimate step towards the exploitation of water as a clean and renewable energy source addresses the energies stored in the low frequencies of liquid flows, which demands flexible solutions to adapt to multiple scenarios, from raindrops to waves, including water moving in pipelines and microdevices. Thus, harvesting low-frequency flows is a young concept compared to solar and wind powers, where triboelectric nanogenerators have been revealed as the most promising relevant actors. However, despite widespread attempts by researchers, the drop energy harvesters' output power is still low, mainly because of the limitations in candidates endowed with ideal triboelectric and wetting properties and also the non-optimal and centimetre-scale device architecture that prevents the conversion of the complete kinetic energy of impinging drops. Herein, we disclose a microscale triboelectric nanogenerator that can harvest a high density of electrical power from drops through a single, submillisecond, long-lasting step. The mechanism relies on an instantaneous electrical capacitance variation owing to the high-speed contact of the drops with the electrodes' active area. We discuss the role of the precharged effect of the triboelectric surface in the time characteristic of the conversion event. The capacitive and microscale structure of the device is endowed with a small form factor that allows for the production of densely packed arrays. The proposed architecture can be adjusted to different liquids and scales and is compatible with a variety of triboelectric surfaces, including flexible, transparent, and thin-film approaches.

**CONTEXT & SCALE**

Water stores immense potential energy through waves, rivers, rain, and tides. Solid-liquid triboelectric nanogenerators (TENG) have emerged to harness kinetic energy from moving or falling drops. In Europe, where prolonged rainy forecasts are common, such energy harvesting will be vital to accomplish the decarbonization agenda. Beyond rain panels, these TENG show promise for powering liquid sensors and microfluidic devices. Here, we present the Trecxel, the first compact TENG to harvest high electrical power density from a single water drop through a one-step process. The size of the Trecxel can be adapted to the mean droplet size, allowing harvesting from a single droplet by Trecxel arrays and enhancing energy density conversion. Trecxel implementation of rigid, flexible, and transparent configurations enables power record values and prompt compatibility with solar panels and glazing. The thin film layout opens the path toward fabrication through low environmental impact and CMOS production.

**INTRODUCTION**

The generation of electrical energy from low-frequency water movement is an emerging concept compared to wind turbines and solar panels.[1–6] Thus, a narrow variety of energy scavenger architectures have been developed during the last decade to harvest from drops and liquid movements available in nature, such as ocean waves, river flows, and rainfall, to produce electric energy as an alternative to the centenary large-scale hydraulic power generation.[7–10] These harvesters work following the electrokinetic effect or as solid-solid, solid-liquid, or liquid-liquid (slippery induced porous surface) triboelectric nanogenerators (TENGs).[7,11–15] Interestingly, most examples for drop-TENGs (D-TENGs) share a common two-step energy generation process that brings two major problems: at first, a falling water drop impacts the charged dielectric (triboelectric) interface and loses parts of the kinetic energy to then slips down towards an exposed electrode or slips over buried electrode for electricity generation.[6,14,16–18] Following this, several novel designs of D-TENGs have been devised. However, attaining substantial instantaneous power output remains a formidable challenge. In 2022, a ground-breaking advancement in the field emerged by introducing a next-generation D-TENG featuring a "transistor-like" structure.[19] This innovation achieved an extraordinary instantaneous power output of 50 W/m$^2$. Also, in this architecture, an essential element is the extended bottom electrode in contact with the bottom side of the triboelectric layer. Very recently, Z. L. Wang and co-workers have reported on the benefits of using arrays of bottom electrodes to increase the responsivity and, therefore, the energy density harvested from multiple droplet impact events.[20] In such a "solar panel-like" D-TENG, a bridge reflux structure is used to independently control each D-TENG, avoiding the movement of charges between the lower electrodes, reducing the inherent capacitance between the bottom electrode and the ground, and by the implementation of a vertical diode structure, leading the charges in a unidirectional path. However, all these proposed architectures still impose relatively large areas for the energy harvesters (typically on the centimetre scale), which become unnecessarily large when dealing with energy coming from falling drops, as in the case of rainfalls, with a typical diameter of a few millimetres. This large form factor results in a significant reduction of the generated power density. In this article, we aim to solve this issue by demonstrating an alternative drop energy harvester with submillimetric dimensions that converts most of the kinetic energy of falling drops and simultaneously improves the output power density by several orders of magnitude.

3D Finite Element Modelling of the drop and the triboelectric device interaction for different liquid conductivities confirms the sudden capacitance variation and charge induction as principal mechanisms for instantaneous power generation. The dimensions, architecture, and compatibility with materials make this triboelectric harvester a game changer in the long-awaited dream of large-scale energy conversion from falling drops, paving the way for rain panels and thin-film-compatible self-powered liquid sensors.

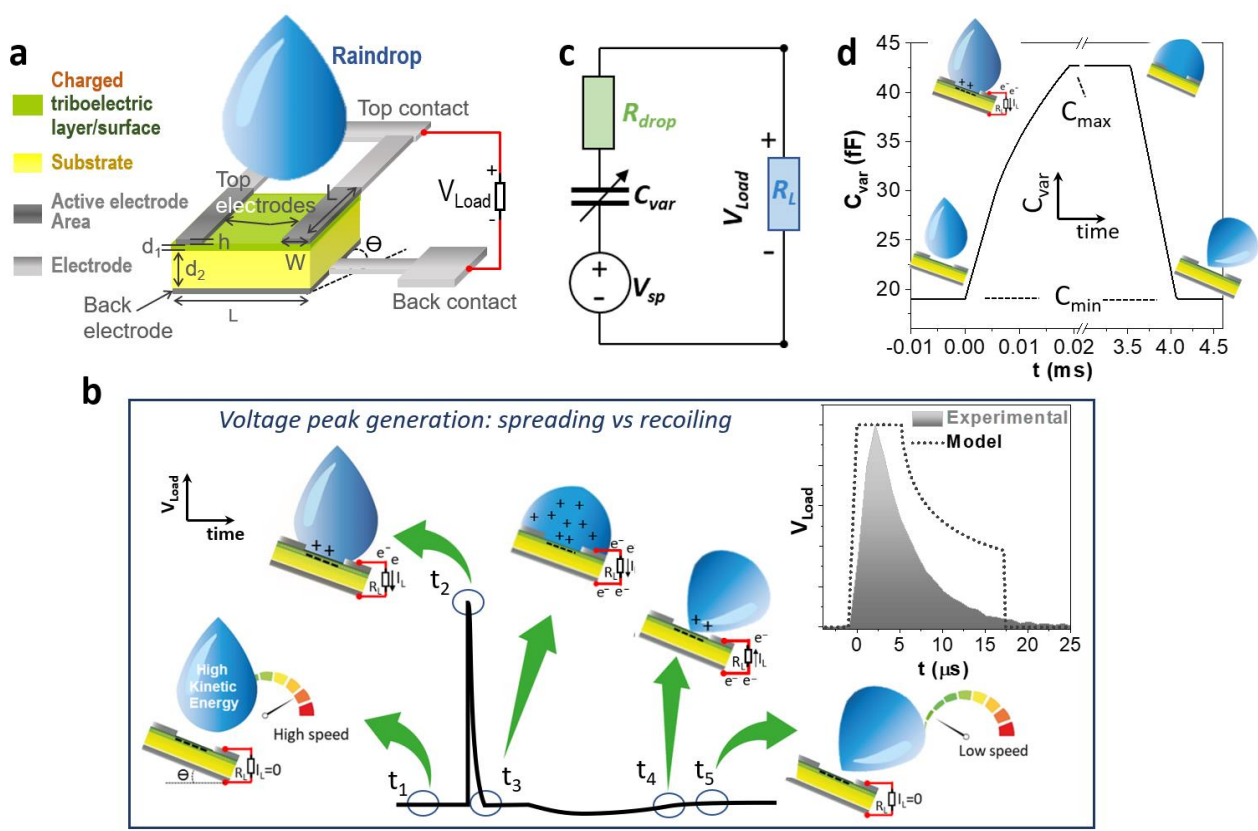

**Figure 1. Trecxel design, working mechanism, and modeling.** a) Schematic of the Trecxel actuated by a raindrop. b) Dynamics of drop-Trecxel interactions and electricity generation timeframe; Inset) Comparison between theoretical (adapted from the analytical model by Pasandideh-Fard et al. described in Supplemental Information 1)[21] and experimental voltage peak generation during spreading in a microscale Trecxel. c) Electrical circuit model that embeds the effect of surface charges ($V_{sp}$), variation of Trecxel capacitance ($C_{var}$), and Drop resistance ($R_{drop}$). d, Simulation of capacitance variation during a complete cycle of electricity generation. See also Video S1 in SI.

**RESULTS AND DISCUSSION**

**Trecxel design and working mechanism**

**Figure 1** presents the design and working mechanism of the proposed triboelectric drop energy harvester, where the capacitive and microscale structure of the device resembles the pixels in a charge-coupled image sensor. Hence, we name it a Triboelectric Pixel or, in short form, Trecxel (Fig.1a). In the first approach, the Trecxel has a top square cross section and an active area of the L×L which includes the back and two top electrodes (dark grey in Fig. 1a). The system is compatible with any type of substrate, rigid or flexible, polymeric or inorganic, consisting of or decorated by a triboelectric layer. The dynamics of the falling drop that impacts the triboelectric surface and the charging state of the surface determine the voltage generated by the Trecxel. In the case of a hydrophobic surface, such dynamics follow a highly asymmetric path, with a very short spreading time (in the range of microseconds) when landing on the surface and much longer recoiling and sliding times (in the order of a few milliseconds).[22–27] To fully elucidate the working mechanism of the device, we have included a detailed description of the spreading, recoiling, and sliding dynamics in the Electronic Supplementary Information 1.

Fig.1b schematically represents the complete cycle of electricity generation caused by the impact of a drop on the active area of the Trecxel. At instant $t_1$), where the output voltage is zero, a high-speed drop is approaching to make a sudden contact with the Trecxel. At $t_2$), the high kinetic energy of the droplet that impacts the surface rapidly deforms it. The generated voltage in this instance is the result of two simultaneous processes: first, charge induction in the drop when it comes in contact with the triboelectric layer, and second, the sudden increase in capacitance of the Trecxel due to the change of effective overlapping area ($A_e$) between the back and top electrodes, which generate a high-power output. Here, the power of such potential difference ($V_{Load}$ as indicated in panel a) of Figure 1) is measured by connecting a resistive load ($R_L$) to the Trecxel. The fast dynamic of the spreading process can be modeled according to the analytical model of Pasandideh-Fard *et al.*,[21,26] which predicts the instantaneous drop diameter in contact with the surface (See SI 1 and video S1). Fig.1b also shows a comparison of the normalized voltage pulse generated by a single Trecxel (for a charged triboelectric surface, grey curve, see experimental details below) and the time derivative of the area (dotted curve, as will be discussed below, the change in the contact area of the drop affects to the delivered voltage) according to the spreading dynamics after the impact of a droplet on a hydrophobic surface. It can be noted that there is perfect agreement in the spreading dynamics regarding the time scale, which predicts a pulse duration of 17.2 µs that accurately matches the observed voltage pulse. Once the active area is covered by the droplet at $t_3$), the power generation is zero as the spreading dynamic stops contributing to the harvesting by this individual Trecxel. After $t_3$), the drop starts to recoil and slide from the hydrophobic Trecxel, which reduces the contact area of the water-solid interface. When the active area of the Trecxel starts to be reduced due to the recoiling and sliding process (between $t_3$ and $t_4$), a negative voltage signal is developed. The detachment process continues up to instance $t_4$) when the overlapping area of the droplet with the active area of the Trecxel is almost zero, which is the last instant that the device delivers a negative voltage. The droplet recoiling and sliding process before rebound is much slower than spreading,[22,24] the output voltage is much smaller (and in most cases barely visible) than the generated at instance $t_2$). Therefore, in this first approach dealing with highly hydrophobic surfaces, we will focus our attention on the positive and instantaneous peak voltage generation during impact and spreading. After a complete separation between the drop and active electrodes in $t_5$), the drop bounces or slides away, and the output current drops to zero. As detailed in SI1, this model mechanism includes not only the effect of the drop dynamics but also the dependency on the electrode dimensions and position with respect to the drop contact.[6]

Fig.1c represents the circuit model of the Trecxel considering the density of surface charges (which develops a surface potential, $V_{sp}$), the variation in capacitance ($C_{var}$), and the effect of the drop conductivity ($R_{drop}$). The model is inspired by previous research works in the field of triboelectric nanogenerators[28,29] and updated to include the dynamic behavior of Trecxel. In this model, $V_{sp} = Q_s/C_0$, which is usually referred to as the surface potential of the dielectric layer, and $Q_s$ is the charge stored on the triboelectric surface (C). These surface charges can be injected by artificial (herein we use an electron beam) or by natural[6,19] means (such as contact with the droplets or by friction with the wind). As detailed below, the injection of electrical charges into the triboelectric layer enhances the instantaneous conversion of energy. In our case, $C_0$ is the equivalent capacitance of the two dielectric layers (substrate and triboelectric layers in Fig.1a, $C_0 = \frac{S\varepsilon_0}{\left(\frac{d_1}{\varepsilon_1}+\frac{d_2}{\varepsilon_2}\right)}$, with thicknesses and relative dielectric constants of $d_1$, $d_2$, and $\varepsilon_1$, $\varepsilon_2$, respectively, with an area of $S=L^2$ and $\varepsilon_0$ is the dielectric permittivity in vacuum). $C_{var}$ is the time-dependent capacitance of Trecxel existing between the two top active electrodes and the back electrode that changes between two values, minimum ($C_{min}$) and maximum ($C_{max}$), as the drop makes contact with Trecxel (Fig.1b and 1d). $Q_{var}$ is the variable charge of the $C_{var}$ with the relation of $V_{var}=Q_{var}/C_{var}$,[30,31] and $R_{drop}$ is proportional to the resistivity of the drop. The simulation results for the given dimensions of the Trecxel area concerning the droplet impact dynamic show that the $C_{var}$ value more than doubles in less than 20 microseconds (Fig.1d and SI 1, see also Fig. S3).

Instantaneous power generation in the Trecxel is highly affected by rapid variations in Trecxel capacitance. $C_{var}$ is the result of the coupling between top and back electrodes and can be defined as $C_{var} = \frac{\varepsilon_0 A_e(t)}{\left(\frac{d_1}{\varepsilon_1}+\frac{d_2}{\varepsilon_2}\right)} = \frac{C_0 A_e(t)}{S}$ where $A_e(t)$ is the effective overlapping area between the top and back electrodes as a function of time. The value of $C_{var}$ is at its minimum before the drop reaches the Trecxel and drop-electrode contact starts to form (Fig.1b and 1d), which is equally formulated as $C_{min} = \frac{C_0 A_e^{min}(t)}{S}$ with $A_e^{min} = 2LW$ (the overlapping area between the top and back electrodes). As contact forms between the drop and the electrodes, $C_{var}$ starts to increase until the drop fully covers the entire surface of Trecxel and fills the entire gap between the two electrodes (Fig.1b and 1d), which makes $C_{max} = \frac{C_0 A_e^{max}(t)}{S}$ with $A_e^{max} = L^2$. Eventually, as the drop slips away from the Trecxel, the drop-electrode contact breaks apart, and $C_{var}$ reduces to $C_{min}$ again. It is worth stressing once again that the reduction in effective area when the droplet recoils and detaches from the surface is a much slower event that produces a reverse current in the load. In our first realization of the Trecxel, this current would be almost negligible compared to most of the previous D-TENGs reported in the literature given the microscale size of the electrodes.[7–14] Additionally, as we will show below, this negative branch will strongly depend on the wetting properties of the surface (wetting contact angle and slippery character of the surface).

To investigate the effect of rapid variations of $A_e(t)$ on the output signal of the Trecxel, the dissipated power over a resistive load is calculated. When a resistive load of $R_L$ is connected to the Trecxel and a water drop hits the surface, a current "$I_L$" will pass through it and the instantaneous power of $P=R_L I_L^2$ is dissipated by the load. Ignoring all sources of energy loss in the circuit, $P$ is equal to the power generated by the Trecxel. The passing current is dependent on $C_{var}$ and $V_{var}$ with the equation of $I_L = C_{var}\frac{dV_{var}}{dt} + V_{var}\frac{dC_{var}}{dt}$.[32,33] Considering the dimensions of the Trecxel in the microscale, together to the two dielectrics used in our first realization (glass substrate and triboelectric PFA sheet), $C_{var}$ magnitude would be in the range of a few femtofarads, therefore by approximation the first term of the equation describing $I_L$ can be ignored compared to the second term producing $I_L = V_{var}\frac{dC_{var}}{dt}$. The power is then calculated as:

$$P(t) = \frac{R_L C_0^2}{S^2} V_{var}^2 \left(\frac{dA_e(t)}{dt}\right)^2 \quad \text{Equation 1}$$

This equation shows that the instantaneous output power is strongly dependent on the change of effective surface area with time and, consequently, with the capacitance variation as $\frac{dC_{var}}{dt} \approx \frac{dA_e(t)}{dt}$. Therefore, the faster the change in effective area in time, the higher the instantaneous output power.

Fig. S2 in SI shows the picture of a first physical realisation of the triboelectric pixel concept (see supplementary information, Video S1 including a slow-motion video showing drop velocity and voltage generated by a single Trecxel). As gathered in Methods, this first Trecxel was produced by a simple shadow-mask processing of sputtered electrodes on a commercially available polymer. In this case, silver was utilized as a conductive material, but the architecture is equally compatible with the application of transparent conducting oxides (TCO) or other transparent solutions with reduced dimensions as presented below. The Trecxel has a square top cross section with dimensions of L=900 µm, W=200 µm, $d_2$=600 µm, $d_1$=50 µm, and h=300 nm. In this example, the substrate is borosilicate glass covered by a triboelectric layer made of perfluoroalkoxy alkane (PFA).

After the microscale assembly of the system, the surface of the triboelectric layer is charged using a beam of 20 kV / 20 µA generated by a field emission electron gun. This procedure provides a high density of stored electric charge on the surface (see Methods) granted by the high electronegativity of fluorine atoms in the PFA polymer.[34,35] Such a Trecxel was used in the experiments to evaluate the energy generated by single drops with a volume in the range of 37 µL impinging on the triboelectric layer from a height in the order of 30 cm. Large droplets have been chosen to simulate a realistic "rain" scenario.[36,37] The graphs of panel a) in **Figure 2** depict the time evolution of the Voltage ($V_{Load}$ vs. time) recorded at different optimal loads at 65 kΩ, 82 kΩ, and 260 kΩ for rain, 60 mS/cm NaCl saltwater, and milli-Q drops.

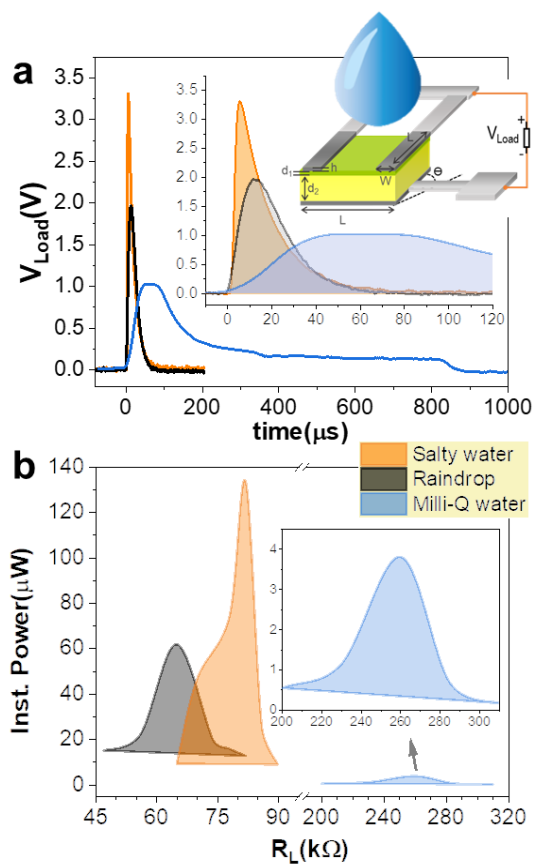

**Figure 2. Characterisation of the Trecxel.** a) Trecxel output voltage (L=900 µm, W=200 µm, $d_2$=600 µm, $d_1$=50 µm, and h=300 nm) in three different optimal resistive loads for different water drops, namely raindrop, salty, and milli-Q water. b) Maximum instantaneous power-load graph showing the output power of the Trecxel for the different water drops and resistive loads.

In all cases, the instantaneous power generation is enough to light a commercially available LED directly connected to the Trecxel, as recorded in **Video S2**. In Fig. 2a, the duration of the voltage pulse is approximately 50 µs for rain and saltwater drops. This transition time is longer than presented in Fig. 1b (see also SI Fig. S3) obtained for a freshly charged surface. We attribute this effect to a lower surface charge in the steady state (see Methods), which makes the initial charge induction in the drop significantly slower. In fact, the transition time is significantly longer (ca. 900 µs) for milli-Q water drops (Fig. 2a). Since the dynamics of the drop impact on the surface should not be modified by the different water drops used here, we attribute such a longer transition time to the low conductivity of the milli-Q water drops, which affect the charge induction time drastically.

The maxima at optimal loads correspond to harvested peak powers (and peak instantaneous output power density) of 134 µW (scaled to 174 W/m$^2$) for salty water, 61.5 µW (76 W/m$^2$) for rainwater, and 3.96 µW (4.9 W/m$^2$) for milli-Q water (considering an active area of 900 × 900 µm$^2$) (Fig. 2b). Such a huge difference in power density for the case of saltwater and rainwater compared to milli-Q water drops is related again the much higher conductivity (60mS/cm, 10mS/cm, 3-6 µS/cm, for salty, rain and milli-Q water, respectively) of the former two compared to the latter (see Finite Element Modeling and simulations). Moreover, the energy provided for a single Trecxel is in the order of 1 nJ (1.36, 1.04 and 0.47 nJ for salty, rain and milli-Q water, respectively). Although this energy is low compared to the potential energy of the droplet (which falls from a height of 30 cm, with a mass of 37 mg, $E_P = m \cdot g \cdot h$ =109 µJ), as it will be shown below, the optimizations developed in this article brings the efficiency to competitive state-of-the-art values for drop-energy harvesting.

Higher conductivity of non-pure water drops allows for rapid exchange of electrical charges between the active electrodes and the drop and faster change of the Trecxel capacitance value. Thus, experiments carried out with nonpolar, nonconductive liquids such as diiodomethane did not produce any power generation. Fig. S4 in SI shows the output characteristic of a control device with a single top and bottom contacts on the millimetre to centimetre scale. This architecture has recently been at the centre of attention by researchers working on the development of solid-liquid TENGs as an early example of instantaneous power generation by this type of energy harvester.[5,6,14,20,38] The experimental methods and materials used for the fabrication and test of the control device are the same as those used for the Trecxel. The instantaneous power obtained with the single-top electrode configuration for salty, rain, and milli-Q water is 5.97 µW ($R_L$=82 kΩ), 1.67 µW ($R_L$=65 kΩ), and 0.06 µW ($R_L$=260 kΩ), respectively, and correspondently to a scaled power density of 9.9 mW/m$^2$, 2.7 mW/m$^2$, and 105 µW/m$^2$ for an active area of 600 mm$^2$. Compared to the above results for the Trecxel architecture, the performance of our system improves the instantaneous power density of the control device in 1.75×10$^4$, 2.8×10$^4$, and 4.6×10$^4$ for salty, rain, and milli-Q water. The calculated ratios indicate more than four orders of magnitude improvement in the output power density when a microscale Trecxel is used for harvesting the energy of a single drop instead of a device with millimetre (top-electrode) and centimetre (bottom electrode and triboelectric layer) scale dimensions. Moreover, the transition time of the three output curves shown in Fig. S4 is in the same range, varying between 2 ms to 5 ms, whereas in the case of the Trecxel (Fig.2a), is in the order of a few tens of microseconds and a few milliseconds, for the conductive (saltwater and rainwater) and milli-Q water drops, respectively. These results and the analysis carried out for different types of salts below show that the Trecxel device is extremely sensitive to the conductivity of the drops, an operational advantage for the development of self-powered water sensors compared to the reported single-top electrode devices.[5,6,14,20,38] Furthermore, these comparative short times for power generation and the reduced dimensions of the device allow for a high-frequency harvesting response to multiple drop events, as shown in **Video S3**, where the peak voltage generated by sequential drops can be followed, depicting high voltage values after successive events.

As mentioned, the impact dynamics cannot explain the differences in the output signal for the different water types employed. Thus, we investigated the underlying mechanisms that govern the relationship between voltage peaks and the conductivity of droplets, employing Finite Element Modeling. Specifically, we conducted simulations using the electrostatics module within the COMSOL Multiphysics package (for detailed methodology, please refer to the SI (Figures **S5** and **S6**) and Methods). These simulations involve a 3D model of the substrate/droplet system (see **Figure 3**). The shape of the drop is described as a 3D expanding "pancake", thereby increasing its contact surface while thinning to keep its volume (37 µL) constant. To describe approximately the effect of the tilted surface of the Trecxel, the centre of the contact surface moves along the surface at a constant speed ($v_{slide} \approx 0.5$ m/s, see SI) during the simulation. In these simulations, the drop and air media are modeled as a single domain and the difference between both is established by a position-dependent relative permittivity $\varepsilon(x, y, z)$ (see Fig. 3 a). Electrostatic simulations are performed under open circuit (OC) conditions, where the bottom electrode is grounded while the top electrodes are left at floating potential. On noncharged surfaces, contact electrification between the pure water drop (milli-Q in the experiments) and the surface may provide almost 90 % of charges by an electron transfer process from water molecules to the polymer surface, while the remaining 10 % relies on ion transfer.[39] On the contrary, the dominant process is ion transfer in the case of charged surfaces.[39] In our case, the PFA surface has previously been negatively charged within the region between the top electrodes. Additionally, we have considered that the rest of the PFA surface is slightly residually charged due to the interaction with the environment and the manipulation of the device. Consequently, the primary mechanism taken into account in our model is the screening of the negative charges on the PFA surface via the transfer of positive ions from the drop to the surface in contact with it (see Fig. S5). Precisely, we have assumed a position-dependent surface charge density, $Q_s(x, y)$, equal to 50 µC/m² within a small rectangular region between the top electrodes (due to the electron-beam charging process) and 0.8 µC/m² in the rest of the PFA surface (related to parasitic/residual charges at the surface) (see Fig. 3a).[28] These values are lower and higher than the surface charge density measured experimentally for the bare PFA in the active area before and just after the charging process (0.3 and 95 µC/m²), respectively. We have assumed these higher/lower charge density values considering that the uncharged/charged PFA surface will naturally charge/discharge with the impact of droplets, as described below for the Trecxel device case. Following our hypothesis, the accumulation of positive charges in the drop as it expands on the PFA surface may originate from two sources: 1) accumulation of Na⁺ ions at the PFA/droplet interface: a fraction of Na⁺ ions, the closest to the double layer, will tend to diffuse to the interface. This diffusion process will occur within the Debye time, $\tau_{Na}$, whose value depends on the concentration of NaCl; 2) accumulation of H⁺ protons, with Debye time $\tau_H$. The time dependence of these accumulated positive charges is assumed to be similar in shape as for the case of a charging (parallel plate) capacitor, i.e., $1 - (1 - f_{Na})e^{-t/\tau_H} - f_{Na}e^{-t/\tau_{Na}}$ (where $f_{Na}$ is the fraction of Na⁺ closest to the interface that participates in the screening of negative charges on the surface). However, the Debye time for protons, $\tau_H$, is around four orders of magnitude smaller than $\tau_{Na}$, thus one can assume $(1 - f_{Na})e^{-t/\tau_H} \approx 0$. Finally, the expression used to quantify the number of positive charges within the drop during contact with the surface is the following:

$$Q(t) = Q_s(x,y)S_c(t)\left(1 - f_{Na}e^{-\frac{t}{\tau_{Na}}}\right), \text{ Equation 2}$$

where $S_c(t)$ is the time-dependent contact area between the drop and the PFA surface (i.e., excluding the electrodes region). This expression guarantees that, for $t \to \infty$, the charges accumulated on the contact surface of the drop will fully compensate for the negative charges on the PFA surface. In this model, the positive charges within the drop are located only at the drop surface in contact with the PFA.

For the simulations shown here, we have considered the spreading and recoiling stages up to a maximum time (see panel 8 in Figure S5 in SI). This is the time when the drop surface contact area returns to zero (see $t_4$ in Fig. 1 b)). Fig. 3b shows the open circuit (OC) Voltage, $V_{OC}$, as a function of time considering the indicated NaCl concentrations diluted in water simulating the three types of water utilized in the experiments. The drop of the OC potential is evidence of the progressive screening of the negative charges on the surface of the PFA. Also note on the close up of the inset of the Fig. 3c how the term $e^{-\frac{t}{\tau_{Na}}}$ allows us to reproduce a delay in the screening as time (and contact surface) progresses. As mentioned above, this delay in the spreading process of the droplet manifests itself as a decrease in the peak voltage and an important increase in the width of the signal. Hence, Fig. 3c shows the voltage peaks obtained from the simulations by considering resistive loads ($R_L$) for salty, raindrop, and milli-Q water, respectively. The inset shows a zoom of the spreading period, where an excellent agreement with the experimental voltage signals can be noted, which have been added for comparison. This fair agreement of the model and the experimental results allows us to analyze the effect of different features on the voltage generation mechanism. On the one hand, we have artificially charged the triboelectric surface between the two top electrodes with an e-beam. However, the voltage peak generation is compatible with the surface homogeneous charging with a lower charge density, mimicking the natural charging of the surface by natural means (e.g., manipulation, friction with air, or drops). In Fig. 3d, we compare the results obtained for these two situations. It is worth noting that the dependence on the type of water is maintained regardless of the surface charging, although the peak intensities decrease for the low-charged Trecxel. In this comparison, we have maintained the optimal load resistance for both charging conditions, which might vary slightly, as we will see experimentally in the following section. On the other hand, the power conversion mechanism proposed for the Trecxel is based on capacitance variation, which makes the system dependent on the triboelectric layer thickness. Simulations in Fig. 3e demonstrated the dependence on the thickness for three different PFA, ranging from 25 to 100 µm, showing that the lower the thickness, the higher the load voltage obtained. This enhancement can be easily understood taking into account that the output voltage can be expressed as (following the same considerations that led to Equation 1): $V(t) = \frac{R_L C_0 V_{var}}{S}\left(\frac{dA_e(t)}{dt}\right)$. Considering that $V_{var}$ and $\frac{dA_e(t)}{dt}$ are not altered by the difference in the thickness, and since $C_0 = \frac{S\varepsilon_0}{\left(\frac{d_1}{\varepsilon_{d1}} + \frac{d_2}{\varepsilon_{d2}}\right)}$, the decrease in $d_1$ explains the increase in the $C_0$ and, consequently, the enhancement of the output voltage.

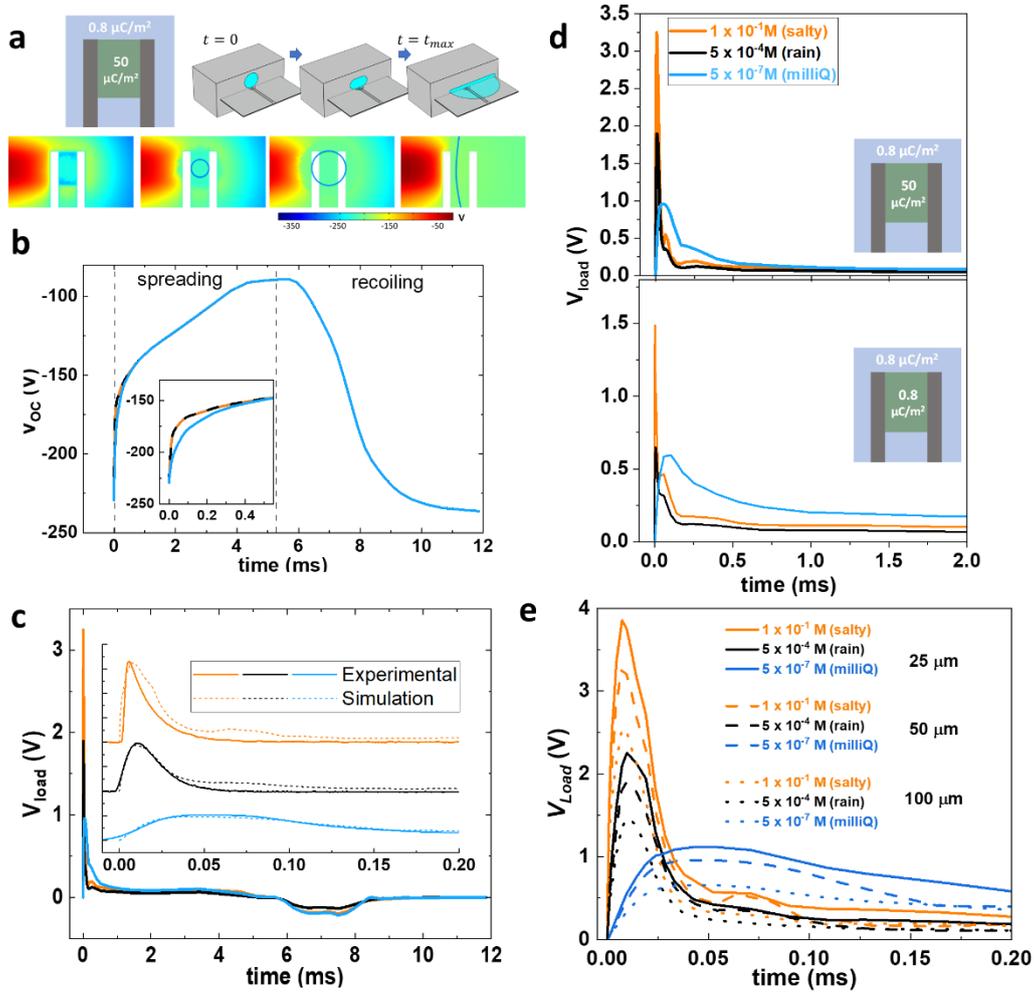

**Figure 3. 3D model of the working mechanism of a single Trecxel.** a) Geometry of the 3D model implemented in COMSOL and surface charge density distribution around the top electrodes, the latter showing the evolution of the drop from t=0 (when the drop just touches the surface at a single point) to the instant when the drop reaches its maximum diameter ($t_{max}$). Specific details are given in the SI and Methods sections. b) Open Circuit voltages as a function of time for three values of NaCl concentration, corresponding to the cases of salty, raindrop, and Milli-Q water cases. c) Load voltages as a function of time for the same three types of drops. The load resistances used in the simulations of each case are the optimal obtained experimentally. The inset shows a zoom in the spreading time region, where the experimental curves have been added for comparison. d) Load voltages as a function of the surface charge density for three types of drops. e) Dependency of the load voltage with the triboelectric layer thickness.

**First steps toward the high-density power conversion by Trecxel arrays**

Referring to Fig. 2, the selected dimensions for the design and fabrication of the Trecxel (such as *L* and *W*) and the number of top electrodes are limited to the applied mask fabrication process. Thus, it is important to stress that the Trecxel energy-harvesting mechanism is compatible with a design to integrate the smallest possible active area so that a single drop covers the whole surface of the device at once upon contact or, ultimately, several Trecxels at the same time. Indeed, to push forward the power delivery of drop energy harvesters and to address their integration in large-area devices, it is critical to demonstrate an approach for energy conversion from multiple events, which in this context involves multiple drops and drop impacts, as it would be required to harvest the energy from rainfalls. A straightforward approach for large-scale energy harvesting from multiple water drops is to form a densely packed n×m array of Trecxels as schematized in **Figure 4** a. Fig. 4b presents the schematic of a 1×2 array of Trecxels packed together with a gap distance of g=200 µm and d= 2 mm. The materials used in the fabrication of the array, experimental conditions, and the dimensions of Trecxels are the same as those used for a single Trecxel in Fig. 2. Fig. 4c shows the output characteristics of each Trecxel connected to a load of $R_L$=65 kΩ, produced by the impact of a raindrop. The temporal dynamics of the $V_{Load}$ for both Trecxels are very similar between them, and to the ones shown in Fig. 2a. When actuated by a single raindrop, the peak power for Trecxel-1 and -2 are measured as 46 µW and 36 µW, respectively. The instantaneous harvested power in total is 82 µW, which is a 33% increase compared to the 61.5 µW power generated by a single Trecxel as shown in Fig.2b. It is worth noting herein that the charging of the triboelectric layer was carried out under the exact same conditions for one and for two Trecxels (i.e. the total charge was distributed on a larger area in the case of the 1 x 2 array, see Methods). This, together with other discharge processes during manipulation (which are likely increased in the case of a 2-Trecxels device), produces the decrease of the surface charge density measured from 96 to 21 µC/m$^2$. Thus, it can be extrapolated that longer exposure times to the electron beam might result in higher output voltages for the arrays.

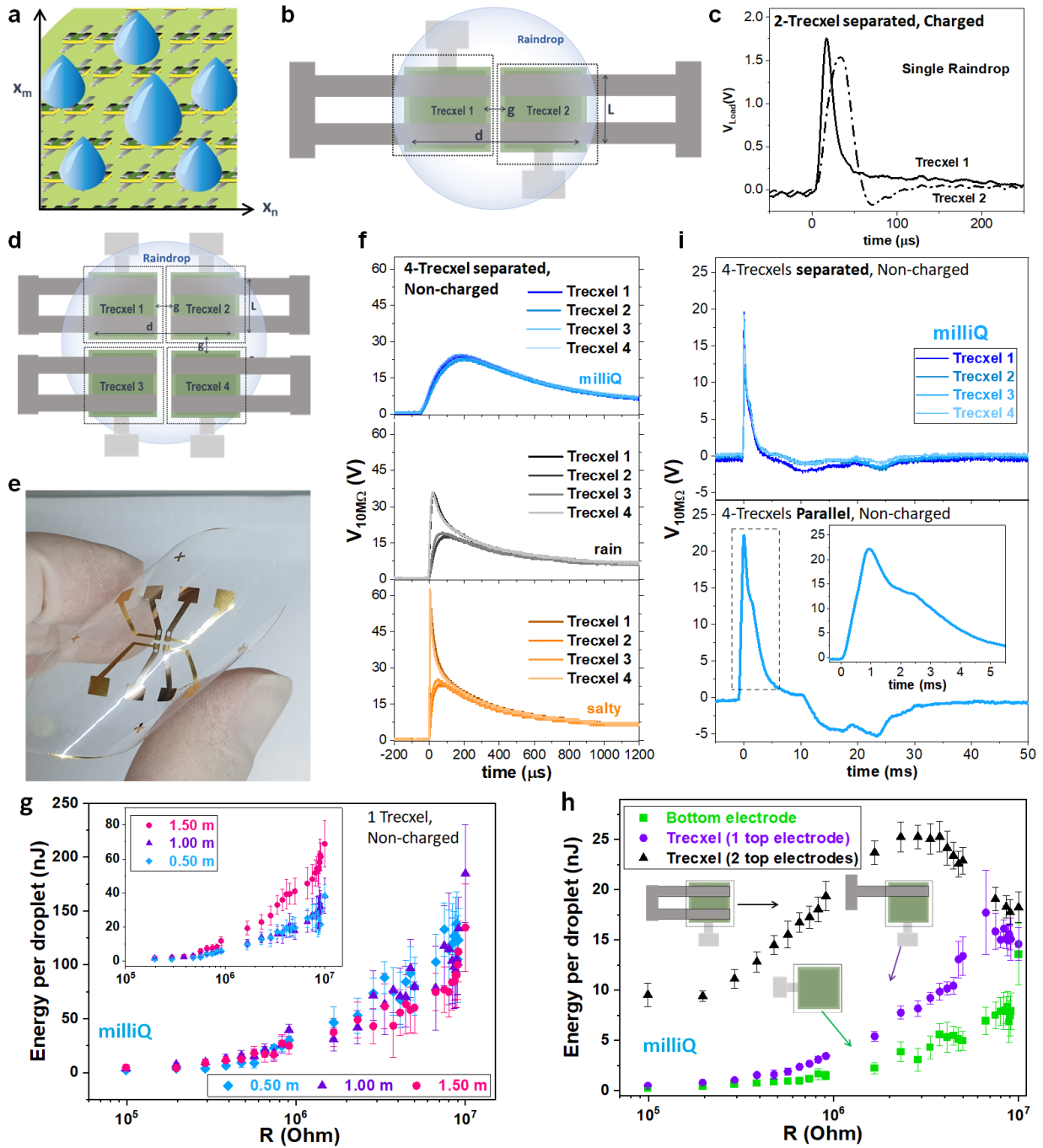

**Figure 4. Trecxels arrays on rigid and flexible configurations.** a) Schematic of a "Rain Panel" that includes an n×m array of Trecxels. b) Schematic of 1×2 array of Trecxels actuated by a single raindrop. c) Output voltage of the two Trecxels connected to a load of 65 kΩ and activated using a raindrop (Trecxels charged and on-supported configuration). d) Schematic of 2×2 array of Trecxels actuated by a single raindrop. e) Optical image of the array used in the experiment in a flexible self-supported configuration. f) Output voltage of the four Trecxels in e, each of them connected to a load of 10 MΩ and activated by milli-Q, rain, and salty droplets of 37 µL. g) Comparison of the energy per droplet obtained as a function of the load for different droplet heights impacting a centered self-supported non-charged Trecxel. The inset shows the response for droplets impacting nearby the Trecxel. h) Dependence of the energy conversion on the number of top electrodes of the devices. i) Comparison of the output voltage of the four different Trecxels, each of them connected to a load of 10 MΩ and activated by a milli-Q droplet (top) and the corresponding output for the four Trecxels connected in parallel (bottom).

Although we cannot experimentally vary the PFA thickness, the architecture of the Trecxel is compatible with removing the glass substrate. For this purpose, bottom/top Au electrodes have been deposited on the back/topside of the triboelectric layer (see Methods). This approach provides a self-supported device that can be easily adapted to different surface topographies and reduces the total thickness of the device. This configuration of the Trecxel allows us to take a step forward, multiply the number of Trecxels, and fabricate them directly on flexible supports. Fig. 4d presents the schematic of the 2 x 2 Trecxel array fabricated by shadow mask on the PFA triboelectric layer. As it is depicted in Fig. 4e, such a self-supported flexible realization permits the adaptation of the arrays to curved surfaces (see also **Video S4**). The 4-Trecxel flexible device was supported on a curved holder with a radius of curvature of ca. 3 cm. In this case, the voltage curves have been measured under load resistance of 10 MΩ. In agreement with previous results, the voltage peaks (Fig. 4f) follow the same trend when comparing milli-Q, rain, and salty water, with the highest instantaneous peak corresponding to the interaction with the first salty drops. An outstanding result of this configuration is that the voltage reaches values over 20 V for milliQ droplets and 60 V for salty droplets, even for non-previously charged triboelectric layers and loads of 10 MΩ, i.e., the voltage output is one order of magnitude larger than for Trecxel supported on glass, as precluded by Fig. 3e, in this case considering that $d_2 = 0$ (i.e., removing the glass thickness).

The curves in Fig. 4f also compare the instant voltage peaks for the different Trecxels in the array, showcasing the effect of the geometry of the impact of the droplet and the contact area between the drop and the triboelectric surface. As can be seen in Video S4, the drop does not contact the four Trecxels simultaneously, impacting earlier with the two upper Trecxels (1 and 4) and then making contact with the lower ones (2 and 3), generating almost half the peak voltage for the lower electrodes. Such a decreased voltage signal for the lower Trecxels is due to the high dependence of the droplet position in the spreading dynamics. As it is concluded from the considerations described in SI 1, 1 mm of deviation in the impact leads to spreading times 1 order of magnitude higher, thus explaining the decreased voltage signal for the lower Trecxels. This feature is much more pronounced for rain and salty than for milli-Q water, revealing the tight relationship between the working mechanism of this drop energy harvester and the conductivity of the drops. This result has opened the question of how the velocity of the droplets affects the energy conversion. Thus, in the proposed mechanism, the velocity defines the drop diameter in contact with the triboelectric surface (Eq. S1), and the variation of the contact area with time is responsible for the power generation (Eq. 1). Fig. 4g shows the energy output obtained for milliQ droplets falling from different heights and impacting on a self-supported Trecxel. The estimation of the energy was carried out from the output voltages of the devices integrating the instantaneous power on the region around the peak by the expression $\int \frac{V(t)^2}{R} dt$ and considering at least 16 drops per point. We have limited the experiments to 1.50 m due to the inherent difficulties of targeting the Trecxel at higher heights. There is a slight variation in the harvested energy for heights between 0.30 and 1.50 m, with maximum values up to 230 nJ obtained for droplets falling from 1.00 m. Contrary to the expected, the trend is not straightforwardly related to increasing heights and, therefore, to the velocities of the droplets. We hypothesize that given the size of the Trecxel and the small distance between the two top electrodes compared with the contact area of the drop, we are not sensitive to the variations yielded by these differences in height. Hence, within the accuracy of our experimental acquisition setup, our system already responds instantaneously upon contact with the drop falling from the lower heights. We have performed a different experiment with droplets impacting with a deviation of the order of the millimeter from the center of the Trecxel (see inset in Fig. 4g). Under these conditions, the energy generated is lower than for the direct impact on the Trecxel as mentioned above, but the expected trend with the drop velocity is more appreciable, with increasing energies for higher falling heights. This result supports our hypothesis regarding implementing a high density of Trecxels of reduced size factor, as the Trecxel arrays would harvest the energy from direct impact and adjacent falling droplets. Moreover, Fig. 4h compares the output energy per droplet for devices formed by the small bottom electrode (voltage output measured between the electrode and ground) and Trecxels with one and two top electrodes to showcase the impact of the number of top electrodes on the energy conversion. The experiments were carried out for non-charged surfaces and milliQ water, demonstrating the improved performance of the devices with top electrodes over the single bottom configuration, in good agreement with the results in reference 6 for extended electrodes, and the increase in harvested energy upon a reduced load resistance for the two top electrodes in comparison with a single top electrode. It is worth stressing that the fabrication by shadow masking implemented in this article hampers the inclusion of additional top electrodes. Nevertheless, the trade-off between the density of devices, number of top electrodes, and exposed triboelectric surface shall be optimized for the different final scenarios, considering the droplets average diameter, frequency, and impact position.

These results further exemplify the importance of producing the Trecxels forming high-density arrays to promote the interaction of the drop with a high number of top electrodes and devices. Thus, the Trecxel in the center of the drop contact area and nearby devices can harvest from a single impact event. Noteworthy, the multiplicity of the triboelectric drop harvesters may increase the complexity of the energy management system, as discussed in reference 20 by Z.L. Wang *et al*. To assess the requirement for individual load connections to each Trecxel, we conducted a comparative analysis of energy conversion between separately and parallelly connected Trecxels (Fig. 4i). Following previous results in panel f), the use case of milli-Q water was selected to ensure that all the connected Trecxels were supplying equal contributions to the total energy. In the case of individually connected Trecxels, the total energy obtained was 0.06 µJ, in comparison to 0.1 µJ for all the Trecxels connected to a single load. This result is not only due to the widening of the positive voltage branch for the parallel Trecxel connection but also to the enhancement of the negative signal corresponding to the recoiling process of the droplet. Therefore, the submillimetric array approach improves the overall energy conversion efficiency and is compatible with a simplified energy management system and device architecture. Moreover, considering the potential energy of the droplet, the efficiency of such parallel Trecxel connections is 0.09%. Although this number is low, the maximum spreading diameter of the droplet is 15.6 mm (see Fig.S2), and assuming a total Trecxel area of 1x1 mm$^2$, a single droplet (V=37 µL, h=30 cm) can activate more than 150-Trecxels. Although the fabrication of such a highly integrated Trecxel architecture will require more sophisticated photolithographic techniques (which will also provide further miniaturization of the Trecxels and maximization of the active area), the efficiency of the Trecxel architecture provided here is highly promising for future drop energy harvesting applications.

**Effect of pre-charging and type of water on the performance of the Trecxels**

The performance of drop energy harvesters is strongly dependent on the triboelectric character of the receiving surface. In Figs. 2 and 4c, we have characterized the response of Trecxels artificially charged with an electron beam. **Figure S7** compares the voltage output from milli-Q droplets for a freshly charged Trecxel with a non-charged one prepared with the curved self-supported configuration. This figure shows how peak intensity and duration depend on the pre-charging state. The voltage achieved for a charged device arises at 42 V with a duration of 1 ms and reduced output (negative voltage) related to the recoil period. Meanwhile, intensity remains half the value (20 V) and duration in the order of 2 ms for the non-charged system. However, it is important to mention that in our experience, the effect of the induced surface charges is only effective during the first measurements, achieving a steady state analog to the non-precharged surface after a variable number of droplets interactions. In the case of salty water, the voltage output becomes equivalent to the pristine uncharged device after several drop events. As shown in **Figure S8**, the maximum power and energy versus load measured with the first salty water droplet depict higher values than the rest of the points, using many droplets. Apart from the measurement using the first drop, the rest of the points measured are very similar for charged and non-charged Trecxels, indicating that the charging effect is lost during the characterization (note that every point in the curve of Fig. S8 requires at least 16 drops). Thus, although the pre-charging of the devices obviously enhances the output voltage, the short duration of the effects generated by direct e-beam irradiation hampers a proper statistical analysis of the system. Alternative methods like corona discharges should be applied to enhance the durability of the surface charges. From now on, we will continue focusing on non pre-treated or naturally charged surfaces. Thus, in previous examples, the surface charging is achieved by friction with air or water, as continuous droplet impinging,[5,6,14,20,38] which induces a high charge density on the triboelectric surface (usually fluorinated polymers) and electrostatically induces opposite charges on the bottom electrode. **Figure 5** shows the evolution of the voltage output upon interaction with different types of droplets, namely milliQ and two types of salty water, NaCl and NH$_4$Cl, and droplets conductivities. The first drop of salty water generates a high peak voltage to then decay (please note the load resistance is fixed at 10 MΩ, i.e., the system is not working at matching impedance). Meanwhile, milliQ droplets present the opposite trend, with the voltage increasing as the number of droplets increases. This experiment shows a slight dependence on the conductivity of the droplet and visualizes the evolution of the voltage upon droplet–surface interaction for the first events.

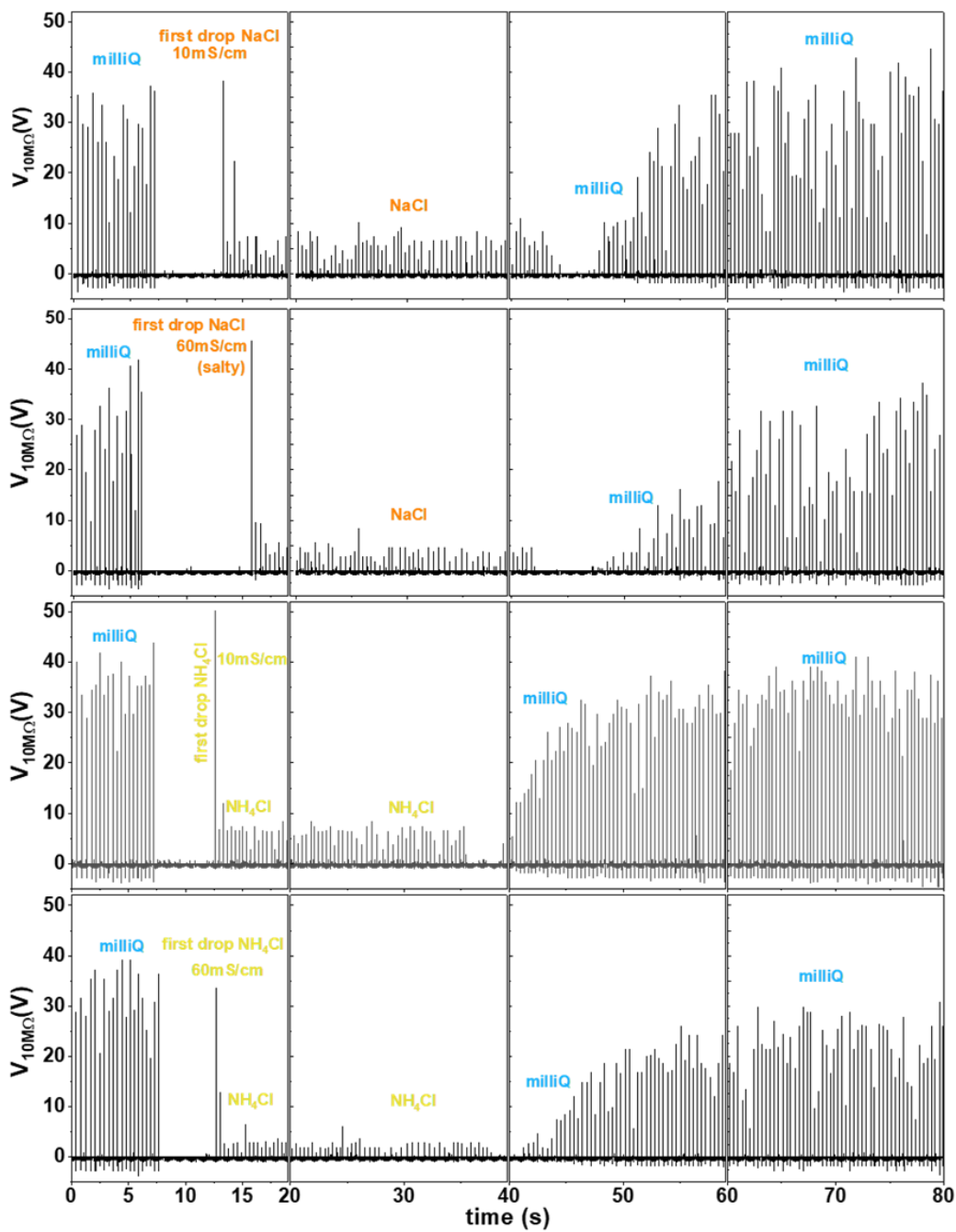

**Figure 5. Voltage output for different cycles of milliQ and salty droplets.** Output voltage of a single-Trecxel for consecutive drops of milliQ and salty droplets for NaCl and NH$_4$Cl with 10 and 60 mS/cm conductivities. Note that NaCl 60 mS/cm corresponds to "salty" droplets in the previous figures.

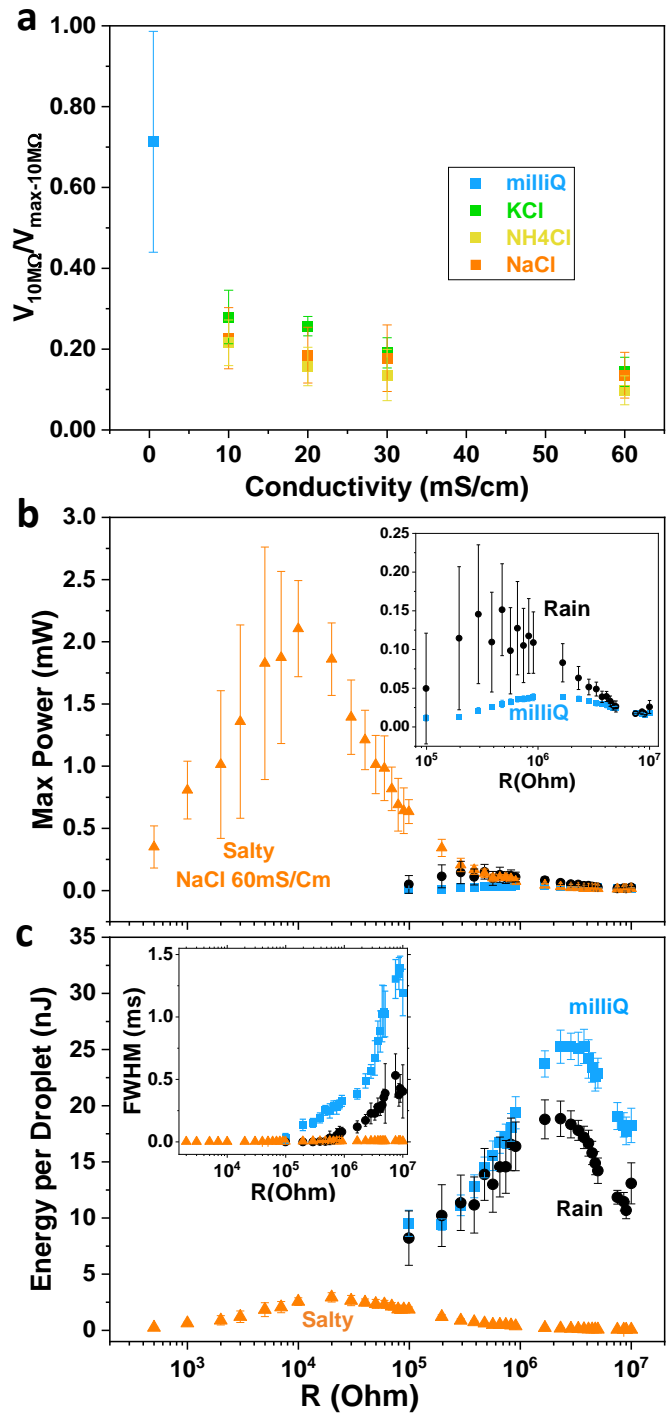

**Figure 6. Performance of non-charged systems for different types of drops.** a) Normalized output voltage dependence with the type of salt and droplet conductivity obtained after consecutive washing with milliQ water for a non-charged Trecxel. Maximum power output (b) and energy per droplet (c) and versus load curves for a self-supported Trecxel on a non-pre-charged triboelectric layer for milliQ, rain, and salty (NaCl 60 mS/cm) drops.

To evaluate the dependence of the Trecxel voltage with the type and conductivity of the droplets, we performed experiments using the same device upon washing for at least two minutes with milliQ droplets and then dripping the drops with different salts and conductivities, removing the first peaks (Figure 6 a). The normalized values indicate a downward trend of the voltage output with the drop conductivity, with the type of cation being a secondary factor. These results have been obtained for a fixed load resistance of 10 MΩ, to statistically evaluate the Trecxel performance, Fig. 6b-c shows the energy and maximum power vs. load curves for non-charged triboelectric surfaces. For each point in the figure, we dropped milliQ water for at least 2 minutes and then carried out statistics in at least 16 drop contact events with a single Trecxel forming part of a 2 x 2 Trecxels configuration. It must be noted that the optimal load is very similar if we consider maximizing the instantaneous power or the system energy. These optimal loads increased from 20.3 kΩ for salty water (i.e., NaCl 60 mS/cm) to 1.95 MΩ for rain and 3 MΩ for milli-Q water. The first key feature of Fig. 6b-c is related to the values of the optimal loads in comparison with the charged single-Trecxel device shown in Fig. 2b. While in the charged single Trecxel device, the optimal loads were in the tens-hundreds of kΩ range (82 kΩ, 65 kΩ, and 260 kΩ, for salty, rain and milli-Q water droplets, respectively), in the self-supported Trecxel device the only value which holds in the same range is the salty water, and the optimal loads for rain and milli-Q water have increased more than one order of magnitude. The results in Fig. S8 show that the pre-charging of the surface is not responsible for this increase of the optimal load, which can be initially understood as the increase in the capacitance for the self-supported configuration used in the self-supported Trecxel devices and has been previously reported to have an impact on the matching load resistance.[40,41] The values of the maximum power output shown in Fig. 6b, and considering an active area of 1x1 mm$^2$, are scaled to 2000, 150, and 40 W/m$^2$, for salty, rain, and milliQ water, which are very high values compared to the literature[6]. Nevertheless, a direct comparison is not straightforward, as contrary to the accurate definition of power efficiency conversion encountered in solar cells, in the case of D-TENGs, there are no standards or conditions for benchmarking the different devices. A remarkable characteristic is that energy is significantly higher for rain and milli-Q water droplets. This feature is caused by an increase in the voltage pulse time, as it is reflected in the Full-Width at Half-Maximum (FWHM) shown in the inset of Fig. 6c. These results are extremely appealing; however, it is important to understand the durability and evolution of the devices upon a high number of interactions with the droplets. With this aim, **Figure 7** gathers the voltage at a load resistance of 10MΩ for non-charged single-Trecxels for long tests with interactions with different types of droplets. Note that we have discarded the first points for the salty droplets in these graphs. The first result is that in the case of milliQ droplets, the voltage output rapidly increases for the first drops (as mentioned in Fig. 5) to continuously grow at a slower pace, at least for the first 60 minutes. For the two type of salty droplets, the trend is opposite, with the voltage output slowly decreasing with the number of droplet impacts. It is worth mentioning that in the case of NaCl (60 mS/cm), the reduction of the voltage output is much more pronounced with time than for the NH$_4$Cl with the same conductivity.

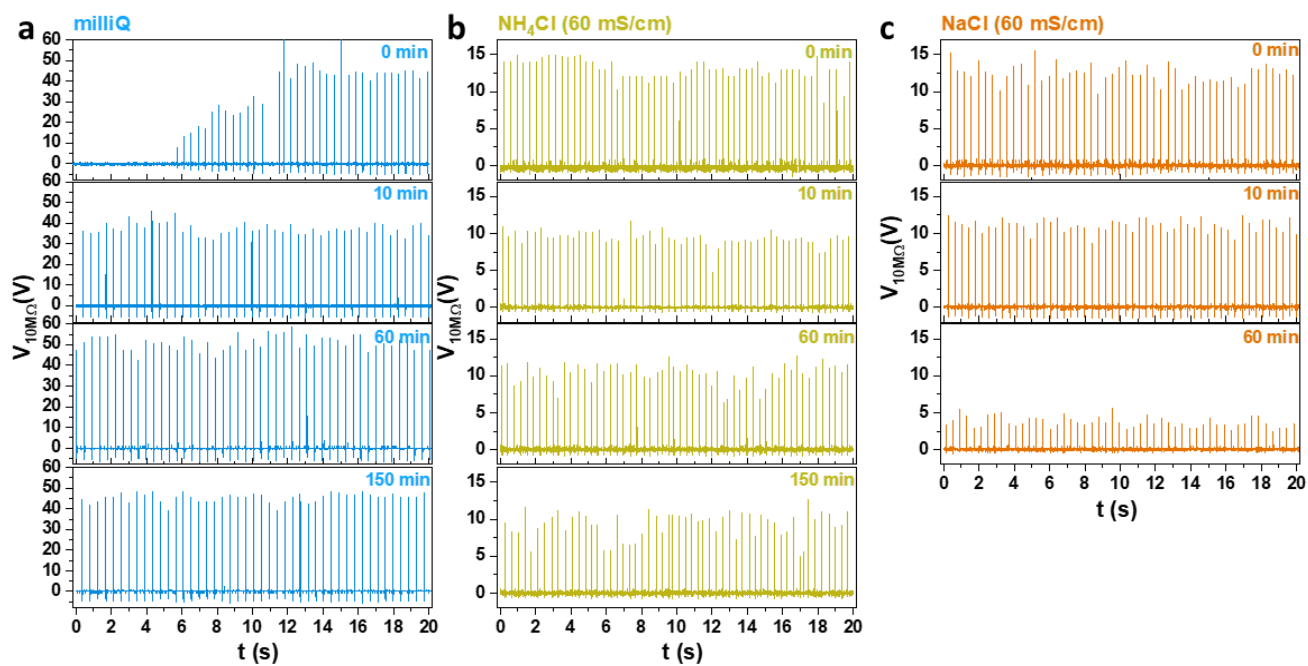

**Figure 7. Voltage output dependence with dropping time.** Output voltage at 10 MΩ of load resistance for different times during dropping milliQ (a), NH$_4$Cl 60 mS/cm (b), and NaCl 60 mS/cm (c) droplets at 2Hz.

Following the results in Fig. 6a showing little dependence on the voltage for these two types of salt, we relate this drop with the more reactive character of the NaCl, which might lead to the faster degradation of the top electrodes. For a fair comparison of the performances, we characterized the energy and power output vs load after ten minutes of dripping each drop type. Thus, the curves in **Figure 8** present a similar trend to the previous result in Fig. 6 but for a much lower peak power output without a clear impedance matching for salty water and similar for milliQ water. This is in good agreement with the trends in voltage output obtained in Fig. 7 between 10 and 60 minutes with time. The energy obtained for milliQ droplets is higher than in Fig. 4, which can be related to slight changes in wettability. These results also highlight that although the system is robust for 18000 interactions with milliQ drops, the interaction with salty drops of high concentration (in this case, 60 mS/cm mimics the sea salty water) eventually saturates the surface (salt precipitation) and affects the durability of the electrodes. Note that in the case of raindrops, the concentration of salts would be smaller (typical conductivity in the order of 6 mS/cm).

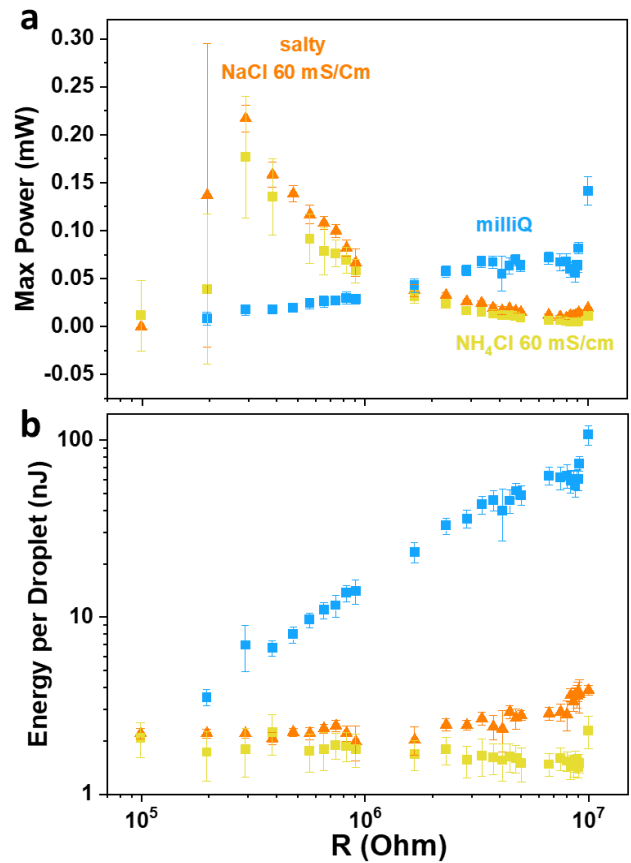

**Figure 8. Performance of non-charged systems for different types of drops for aged devices**. Maximum power output (a) and energy per droplet (b) versus load resistance curves for a self-supported Trecxel on a non-pre-charged triboelectric layer for milliQ and salty (NaCl 60 mS/cm and $NH_4Cl$ 60 mS/cm) drops after dripping for more than 10 minutes. The curves are acquired from large to small resistances.

**Thin film Trecxels**

The eventual realization of the high-density array would provide a way to harvest kinetic energy from the same droplet by many Trecxels at the same time and from recoiling and bouncing droplets by adjacent Trecxels. However, one of the most important challenges to overcome for the fabrication of such an array employing high-yield, short payback time, and industrially available manufacturing techniques is the compatibility of the Trecxels with thin-film (triboelectric) technology, a topic only scarcely explored in the case of solid-solid TENGs.[42,43] Thus, the last part of this article aims to introduce the thin-film approach applying standard vacuum and plasma steps and materials commonly used in the thin-film microelectronic industry. Such an array can be fabricated using advanced high-yield manufacturing techniques such as photolithography technology, roll-to-roll, and 2D printing.[44–46] **Figure 9** a shows an example of a thin film 2 x 2 array prepared with a transparent combination of materials (a similar architecture was developed on Si wafer using gold electrodes deposited by thermal evaporation), namely, transparent conducting oxide (ITO)[47] fabricated by DC pulsed magnetron sputtering and 300 nm $SiO_2$ thin film formed by Plasma Enhanced Chemical Vapor Deposition (PECVD)[48] functionalized with a perfluorinated molecule (PFOTES)[49] working as triboelectric surface. The role of PFOTES is twofold: to increase the WCA of the otherwise hydrophilic SiO$_2$, and to enhance the surface charging of the surface.[49] Even though the PFOTES molecule increases the WCA (see Fig. S1b), the hydrophilic character of the surface is not totally suppressed, thus having consequences in the voltage generated by the drop, as it is detailed below. Fig.9 b and c showcase the voltage outputs corresponding to different device configurations (as labeled) for single and 2 x 2 arrays correspondently (see Methods for fabrication details). Fig. 9c shows the voltage pulses recorded for every individual Trecxel prepared on the transparent fused silica substrate with the triboelectric layer sandwiched between the top-bottom electrodes. The first characteristic of the voltage peaks is the absence of a positive branch, and the appearance of a major negative peak in the output of all the Trecxels in the array. The reason for both effects is the much higher hydrophobicity of the PFA surface compared to the oxide surface, with values of water contact angles using rainwater of 100 and 50º, respectively (see Fig. S1). The partially hydrophilic character of the PFOTES-grafted SiO$_2$ surface produces its wetting during the first drop interaction. Consequently, the following droplet impact cannot produce an increase in the Active area of the Trecxel ($A_e$), which is fully wetted before the droplet arrives. However, the impacting droplet can induce the formation of a trapped air layer, which is displaced from the center to the radial directions.[50] This air layer produces a sudden initial decrease in the $A_e$, which at the same time, produces the observed negative voltage peak. A comparison of the outputs of Fig.9 and Fig. 2 and 4 reveals two major differences. The first is the substantial difference between the transition times of the voltage peaks (note the different time scales on the x-axis), which are in the range between 50 μs to 1 ms for the Trecxel fabricated with a PFA layer as the triboelectric layer and in the order of tens of milliseconds for the thin film device (Fig. 9b). The device shows a very low instantaneous power of 0.02 μW and the total energy generated is around 0.6 nJ per Trecxel (note that in this case the impedance matching has not been characterized). Despite these poor values, the time increase observed is very promising for energy harvesting applications. Therefore a similar supported single Trecxel device was developed on intrinsic Si(100) wafer. In this case, the device was connected to the optimal load of $R_L$= 47 kΩ for characterization. Again, highly broad voltage pulses are shown in Fig. 9c, but in this case, a minor positive voltage was observed, indicating a slightly different wetting behavior, likely connected to the electrodes rather than to the PFOTES-grafted SiO$_2$ surface, that was similar in both cases. In this case, a maximum instantaneous power of 5.7 μW and an energy of 59 nJ are calculated. This value represents an efficiency of 0.06%, a highly promising value considering that a single droplet of these characteristics can activate around 150-200 Trecxels in a single impact, without considering rebounds or slips.

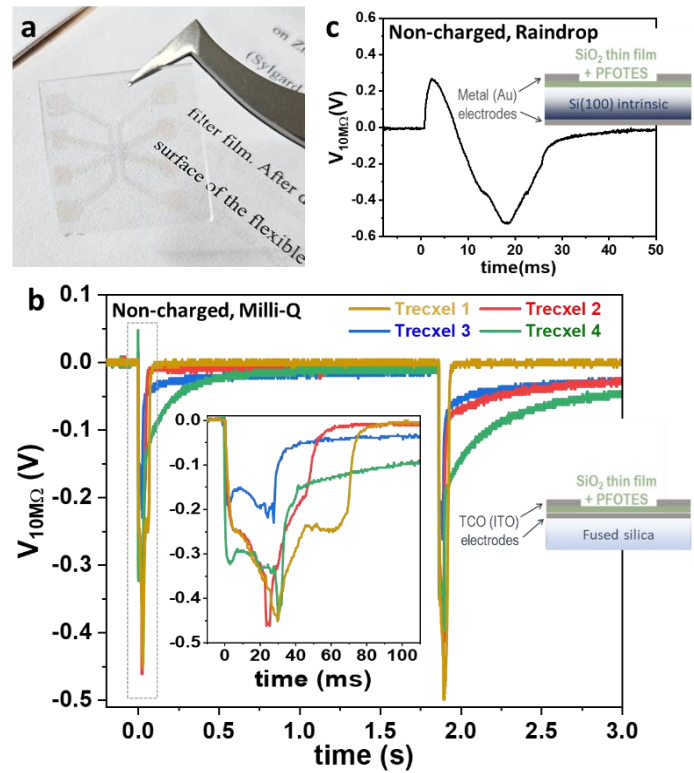

**Figure 9. Thin film Trecxel configuration.** a) Photograph of the thin film devices for 4 Trecxels for ITO (transparent) electrodes. b) Voltage output obtained for the impact of multiple milli-Q droplets on 4 different Trecxels fabricated on fused silica. c) Voltage output obtained for the impact of a raindrop on a single Trecxel fabricated on intrinsic Si.

## CONCLUSIONS

A novel solid-liquid triboelectric nanogenerator architecture compatible with submillimetric dimensions (Trecxel) has been demonstrated. The small form factor allows for two critical features: on the one hand, the instantaneous power conversion from falling drops (down to microsecond range), and on the other hand, for the assembly of nanogenerator arrays to harvest from a single impact event from several TENGs simultaneously. The Trecxel is straightforwardly compatible with flexible, self-supported, and thin-film configurations. We have analyzed the power conversion mechanism, revealing the role of the sudden capacitance variation, given the droplet interaction with the top submillimetric electrodes, the surface contact angle and slippery properties, and the chemical characteristics and conductivity of the drops. Also, the pre-charging of the triboelectric layer by artificial means, i.e., an electro gun, has been revealed to increase the voltage output for the first interactions of droplets and the surface. Although this pre-treatment does not generate durable surface charges, alternatives like corona discharge activation of the triboelectric layer are compatible with this architecture. The dependence of the device performance on the size and characteristics of the system have been evaluated, resulting in enhanced energy conversion for two top electrodes, in comparison with one, and for thinner triboelectric layers and a smaller size factor. Au electrodes deposited by thermal evaporation have yielded lifecycles of the nanogenerators beyond 18000 milliQ droplet impacts. Conducting droplets generates a higher maximum power conversion at lower loads. However, the stability of the energy harvester is compromised after long testing periods with highly conducting drops, likely due to the degradation of the top electrodes or the formation of precipitates among them. Implementing Trecxels arrays allows for an enhancement in the power conversion up to 33% (charged) / 67% (self-supported, non-charged) comparing two / four Trecxels with a single-TENG configuration. Moreover, working on thin-film configuration has yielded energy conversion ca. 60 nJ per droplet impact. These results indicate that the Trecxel design allows for power generation even for non-hydrophobic slippery surfaces, albeit on a different time scale. They also demonstrate the facile implementation of the triboelectric pixels to form high-density arrays with *ad hoc* dimensions to respond to determined drop volumes and spreading/recoiling dynamics and the potential to promote the specialized academia in superwettable and tunable wetting surfaces into this exciting topic. More importantly, the compatibility with thin-film technology approaches prompts the application of down-micron-sized packed arrays to harvest the energy of high-frequency events and strong rainfalls.

High-density arrays of drop energy harvesters hold promise as practical solutions for the realization of rain panels. Moreover, the compact size of Trecxels and their adaptable configuration make the triboelectric architecture especially appealing for harvesting energy from fog and droplets in maritime environments. As exemplified by the phenomenology observed, the droplets chemical composition and conductivity influence the nanogenerators voltage and power output, a characteristic shared with other D-TENG approaches documented in the literature. [5,6,14,20,38] The optimization of the Trecxels for the different scenarios requires adaptation of the size, number of top electrodes, wettability, and density of the arrays. Power management systems capable of responding to the high-density events of microenergy conversion need to be developed alongside.

**EXPERIMENTAL PROCEDURES**

Acetone (VWR, 99.9%), Ethanol (VWR, 97%), PFA films (Goodfellow, 50 μm thickness), laboratory cover glass (borosilicate by Labbox), Kapton films (Goodfellow), silicon wafers (Topsil, 4 inches, (100), resistivity (Ω.cm)>10000), 500 μm thickness), pure water (Merck Millipore, Milli-Q water).

*Fabrication of the Trecxel*

Different sets of substrates and triboelectric layers were used for the fabrication of the Trecxels. For fabrication of the glass-PFA Trecxel (Fig.2a), a cover-glass with dimensions of 20 mm × 20 mm × 600 μm was cleaned and washed on both sides with ethanol and acetone, and a film of PFA was cut into a square of 20 mm × 20 mm and then washed and cleaned with milli-Q water. The PFA sheet is taped with transparent double-sided tape on the top surface of the glass substrate. For the PFA self-supported flexible devices, a PFA sheet was used as the substrate, directly supporting the electrodes on both sides. For the case of the thin film compatible Trecxel, fabrication started from a double side polished intrinsic silicon wafer or fused silica slips, which were cut into a 20 mm × 20 mm square and ultrasonically cleaned with ethanol and then acetone. Then, a thin layer of silicon oxide is deposited by Plasma Enhanced Chemical Vapor Deposition (PECVD) at room temperature following the protocol shown elsewhere.[48] Chlorotrimethylsilane from Sigma Aldrich was used as the delivered Si precursor. The oxygen and Si precursor were supplied in the reactor using a mass flow controller. The base pressure of the chamber was lower than $2·10^{-5}$ mbar, and the total pressure during deposition was around $10^{-2}$ mbar. The plasma was generated in a 2.45GHz microwave Electron-Cyclotron Resonance (ECR) SLAN-II operating at 800 W, and the deposition was carried out in the downstream region of the reactor. The thin film thickness was settled at 300 nm. After the deposition, chemical derivatization of the surface was carried out by exposing the sample at 80ºC to 1H,1H,2H-Perfluorooctyltrietoxisylane (PFOTES) 98% vapor on a static low vacuum (obtained with a rotatory pump) at least for 2 hours.[51] Thus, in this case, the silicon plays the role of the substrate, and the perfluorinated silicon oxide plays the role of the triboelectric layer.

The bottom and top electrodes were produced by masking and metal sputtering. First, non-adhesive Kapton films or stainless steel sheets were patterned with the geometry of the top and back electrodes through a laser ablation process (Explorer®One, 355 nm wavelength). Silver electrodes were deposited over the triboelectric layers by DC sputtering through both aligned top and bottom masks. This deposition was carried out in an argon atmosphere with a total pressure of $1.6·10^{-1}$ mbar and a voltage of 600 V for plasma generation. The base pressure of the chamber was lower than $10^{-3}$ mbar. The deposition time was 2 hours for each electrode to get fully conductive layers with a thickness of ca. 300 nm. Gold-titanium electrodes were deposited by evaporation. In a vacuum chamber (total pressure lower than $10^{-5}$ mbar), a thin film of 2-3 nm titanium (Ti) was deposited directly by evaporation on the samples through the shadow mask. Just after that deposition, gold (Au) evaporation was performed from another evaporator in the same chamber. Finally, the 80 nm gold thin film was deposited on top of the previous titanium layer through the same shadow mask. ITO thin films as bottom and top electrodes were deposited by rf magnetron sputtering, using a 2" ITO target (Indium-Tin oxide 90-10wt%), 99.99% pure from Testbourne. The base pressure of the chamber before deposition was lower than $5·10^{-6}$ mbar, and a working pressure of $5·10^{-3}$ mbar of pure Ar was employed. The substrates were heated up to 350 ºC, which was kept constant during deposition. The power applied to the target was 50W, and the distance between the target and the substrates was 10 cm.

*Setup for measurements and characterizations*

For all experiments but for the characterization at different heights in Fig. 4, water drops are released from a distance between 30 and 50 cm by a 5 mL polyethylene pipette onto the Trecxel surface. The pipette dispenses water drops with a volume of 37 µL. The pipette is gripped and rigidly fixed by a laboratory stand and clamps. Trecxel is fixed onto a tilted flatbed, which makes a 30º tilting angle with the horizon. The output voltage of the Trecxel is measured using a Tektronix 3024 oscilloscope. A high impedance probe with 10 MΩ of internal resistance is used to measure the voltage across different loads.[52] Measurements were performed in a lab environment with humidity of 65% and 20º C.

Rainwater was collected on the 26$^{th}$ and 27$^{th}$ of April 2021 in Seville, Spain. Salty water was prepared as a NaCl solution in milli-Q water with a concentration of 0.585 g of NaCl per 100 mL of water to yield a conductivity of 60 mS/cm. For the experiments in Fig. 4-9, the load impedances were changed using a resistance box, and the Trecxel output voltage was measured using a PicoScope oscilloscope. Energy and peak power conversion vs load curves were characterized from high to low load resistance values. The rainwater was collected on 18$^{th}$ of May 2023 in Seville, Spain, and its conductivity was around 600 µS/cm. Sodium chloride, potassium chloride, and ammonium chloride from Sigma Aldrich were used to obtain the different solutions we used to test the Trecxels under different liquid conductivity conditions. To prepare the high-conductivity solutions, we mixed the salt with milliQ water to control the weight of the salt, volume of the solution, and conductivity. Water conductivity measures were performed using a conductivity probe 731-ISM Mettler Toledo InLab as the conductivity module for the Mettler Toledo SevenExcellence multiparameter.

*Measurement of Water Contact Angle*

A water contact angle (WCA) measurement system (OCA20 goniometer from DataPhysics) was used to measure and compare the hydrophobicity of the PFA and SiO$_2$ surfaces with respect to raindrops (Extended Data Fig. 1). The WCA system consists of a peristaltic pump (Perimax 12/4-SM from Spetec) to control the water flow and a set of calibrated micropipette tips that ensure a certain size of the falling droplets.

*Surface Charging Protocol*

The triboelectric layers of the Trecxels were charged using an electron beam in a Scanning Electron Microscope. All samples were exposed for 10 minutes to a beam acceleration voltage and current settled at 20 kV and 20 µA, respectively, over the active area of the Trecxel (0.9*0.9 mm$^2$). Recently charged surfaces present abrupt voltage curves. To have a more realistic scenario, the surfaces (otherwise stated) were measured several hours after charging and impacting with more than 100 droplets to reach a steady state. Surface electrostatic charges were characterized by employing a non-contacting electrostatic voltmeter (ESVM 1000, Wolfgang Warmbier GmbH) with an active area of 40 mm$^2$, measured at 10 mm from the surface. Before the charging procedure, the measured potential is (mean values) -33 V, corresponding to a surface charge density of $\sigma_{uncharged}$=0.3 µC/m$^2$. After the charging protocol for the device with a single Trecxel, the potential is increased to -270 V. If we assume that the charge is located in the active area defined by the beam used in the microscope of 0.9*0.9 mm$^2$, and considering that the surface charge density out of this square is equal to the charge density measured before the charging protocol, the surface charge density of the active area is $\sigma_{charged}$=96 µC/m$^2$. For the 2 Trecxels device, the measured potential before and after charging protocol result in an increment from -61 V to -177 V. These values correspond to a surface charge density before (also outside the active area) and after the charging protocol (in the active area, which now is 0.9*2 mm$^2$) of σ$_{uncharged}$=0.5 µC/m$^2$ and σ$_{charged}$=21 µC/m$^2$. For the 4 Trecxels PFA flexible device, the measured potential before charging was -22 V, and it increased to -750 V, which corresponds to a surface charge density before and after the charging protocol (in the active area, which now is 4*0.9*0.9 mm$^2$) of σ$_{uncharged}$=0.2 µC/m$^2$ and σ$_{charged}$=61 µC/m$^2$.

*Calculation of average and instantaneous power densities and energy per droplet.*

Assuming that $V_{Load}$ is the measured voltage across a resistive load of $R_L$, the instantaneous power is calculated as, $P_L(t) = V_L^2(t)/R_L$, where the $P_L^{max}$ is the maximum value of $P_L(t)$. Assuming that the active area of the drop energy harvester is $A$, the maximum power density is defined as: $P_L^{max}/A$.

Proprietary software (NanoDataLyzer) developed by the authors facilitated statistical analysis of the maximum power conversion and energy per droplet. This software is available in open access in Zenodo.[53] Each point in the graphs in Figs. 4 to 8 has been obtained upon an average of at least 16 droplets impacts.

*COMSOL Modeling of the Trecxel*

The V$_{Load}$ was obtained from $V_{OC}$ and $C_{var}$ by numerical simulations using the electrical circuit module of COMSOL package. We have found that, in order to reproduce the experimental load voltages, it is necessary to consider a maximum capacitance at least one order of magnitude larger than the one obtained theoretically by geometric considerations. The origin of this mismatch is unclear, but previous works[54] point out to the imperfections and roughness of the dielectric/electrode surfaces as the main origin of the increase in the experimentally observed capacitance with respect to what can be estimated theoretically from the geometry. It is important to note that the dependence of this increase in $C_{var}$ on roughness is non-linear, so it seems reasonable to find increases of up to two orders of magnitude due to roughness and imperfections, electrodes edge effects, etc., on the surface in combination with the dielectric characteristics of the water drop. Assuming these features as the origin of the capacitance mismatch, we have multiplied our theoretical capacitance by a factor $N_c$. We also considered the overlap between the drop+top electrodes and all the bottom electrodes in the simulation of the time-dependent capacity (see Figure S6).

Additionally, we have included the possibility of a "parasitic capacitance", $C_p$, in the equivalent TENG circuit (see **Figure S9**). According to K. Dai *et al*.[55] these parasitic capacitance are unavoidable and usually range between few pF and hundreds pF. These are assumed to come from a combination of sources: external metal wires, capacitances in the measurement equipment, external circuit boards, etc. Since the parasitic capacitance degrades the average power for all values of the load resistor, its main effect is to reduce the optimal load but also reducing the power. In addition to the electrodes, the modeled Trexcel consists of three layers PFA/Kapton/borosilicate with thicknesses 25, 50 or 100µm/70µm/130µm and relative permittivities 2.1/4/4.8.

## AUTHOR CONTRIBUTIONS

XGC fabricated all the devices with support from FNG and VG. AG carried out the characterization of the liquid-solid TENG for single droplet events of precharged devices. XGC has carried out the statistical analysis and characterization of the devices. XGC and JRSV were responsible for the simulations and the drop dynamics versus power analysis. AG is responsible for Eq. 1. JB carried out the simulations and modeling. CLS assisted in the contact angle and device characterization. JRSV, AB, and AnaB did the conceptualisation. AG and AnaB wrote the original draft of the article. All authors contributed to the final version of the manuscript.

## DECLARATION OF INTERESTS

Dr. Ali Ghaffarinejad current affiliation is Hanzle UAS, the Netherlands. The content of this article is related to two patent applications: 1) A. Ghaffarinejad, X. García-Casas, A. Barranco, J. R. Sanchez-Valencia., A. Borras. WO 2023/025965 A1. 2) X. García-Casas, A. Barranco, J. R. Sanchez-Valencia, A. Ghaffarinejad, A. Borras. PCT/EP2024/053180.

## ACKNOWLEDGEMENTS

The authors thank the projects PID2022-143120OB-I00, PCI2024-153451 and TED2021-130916B-I00 funded by MCIN/AEI/10.13039/501100011033 and by "ERDF (FEDER)" A way of making Europe, Fondos NextgenerationEU and Plan de Recuperación, Transformación y Resiliencia". C.L.S. thanks the University of Seville through the VI PPIT-US and "Ramon y Cajal" program funded by MCIN/AEI/10.13039/501100011033. JRSV acknowledges project ANGSTROM was selected in the Joint Transnational Call 2023 of M-ERA.NET 3, an EU-funded network of about 49 funding organizations (Horizon 2020 grant agreement No 958174). XGC acknowledges the FPU program under the grant number FPU19/01864. FNG acknowledges the "VII Plan Propio de Investigación y Transferencia" of the Universidad de Sevilla. The project leading to this article has received funding from the EU H2020 program under grant agreement 851929 (ERC Starting Grant 3DScavengers)

# Supplemental Information

# Triboelectric Pixels as building blocks for microscale and large area integration of drop energy harvesters

Ali Ghaffarinejad,[+a] Xabier García-Casas,[+a*] Fernando Núñez-Gálvez,[a,b] Jorge Budagosky,[a,b] Vanda Godinho,[a] Carmen López-Santos,[a,b] Juan Ramón Sánchez-Valencia,[a] Ángel Barranco,[a] Ana Borrás[a*]

[+]These two authors have contributed equally to the work

a) Nanotechnology on Surfaces and Plasma Laboratory, Consejo Superior de Investigaciones Científicas (CSIC), Materials Science Institute of Seville (CSIC-US). c/ Américo Vespucio 49, 41092, Seville (Spain)
b) Departamento de Física Aplicada I, Universidad de Sevilla, C/ Virgen de Africa 7, 41011, Seville (Spain)

*anaisabel.borras@icmse.csic.es; xabier.garcia@icmse.csic.es

**Supplementary Information S1. Droplet impact dynamic on hydrophobic surfaces.**

PFA and perfluorinated $SiO_2$ surfaces depict different water contact angles, as presented in Fig. S1, in concordance with their different chemical and surface roughness properties.

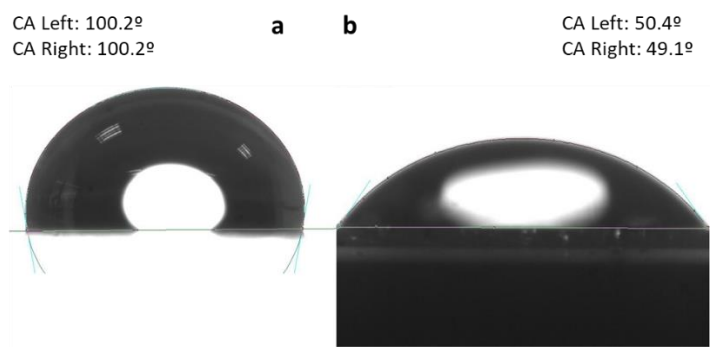

**Fig. S1 | Contact angle generated between a raindrop and different surfaces.** Static contact angle between the drop and PFA surface **(a)** and the PFOTES functionalized silicon oxide surface **(b)**.

Surface wetting properties, drop volume, and landing velocity determine the dynamics of a bouncing droplet that, on a hydrophobic surface as the PFA, follows a highly asymmetric path, with a very short spreading time and a significantly longer recoiling time.[1-5] In this section, we aim to analyze how the droplet spreading and recoiling dynamics affect the power generation by the Trecxel looking for the elucidation of the power generation mechanism. In the calculations below, we will refer to the PFA case for a tilting angle of 30º to ensure the proper sliding of the drops (with 37 µL of volume and falling from a distance of ca. 30 cm). It is worth stressing that although the settled tilting indeed may modify the dynamic of a bouncing droplet, it will be only taken into account during the recoiling process, due to the very fast dynamics of the spreading.

*Supplementary Information 1.1: Dynamics for a spreading droplet*

The fast-spreading dynamic of a drop onto a surface is usually approximated by the analytical model of Pasandideh-Fard *et al.*[6] which predicts the instantaneous diameter as:

$$\frac{D(t)}{D_0} = \left(\frac{D_{MAX}}{D_0}\right)\sqrt{\frac{3vt}{8D_0}} \quad \text{(Equation S1)}$$

where $D(t)$ and $D_{MAX}$ are correspondently the drop diameter as a function of time ($t$) and the maximum diameter of the droplet in contact with the surface, $D_0$ is the diameter of the droplet before the impact and $v$ is the speed of the droplet at the impact. We have used droplets with volumes of 37 µL ($D_0$=4.14 mm) thrown from a height of 30 cm, providing a speed of approximately $v$=2.4 m/s (see Video S1). Under these conditions, for our hydrophobic surfaces with static water contact angles of 100º (see Fig.S1), we have measured a $D_{MAX}/D_0$=3.7 ($D_{MAX}$=15.6 mm, Fig.S2). Herein, we have used such large droplet sizes and high impact speeds to have a more realistic rain scenario.

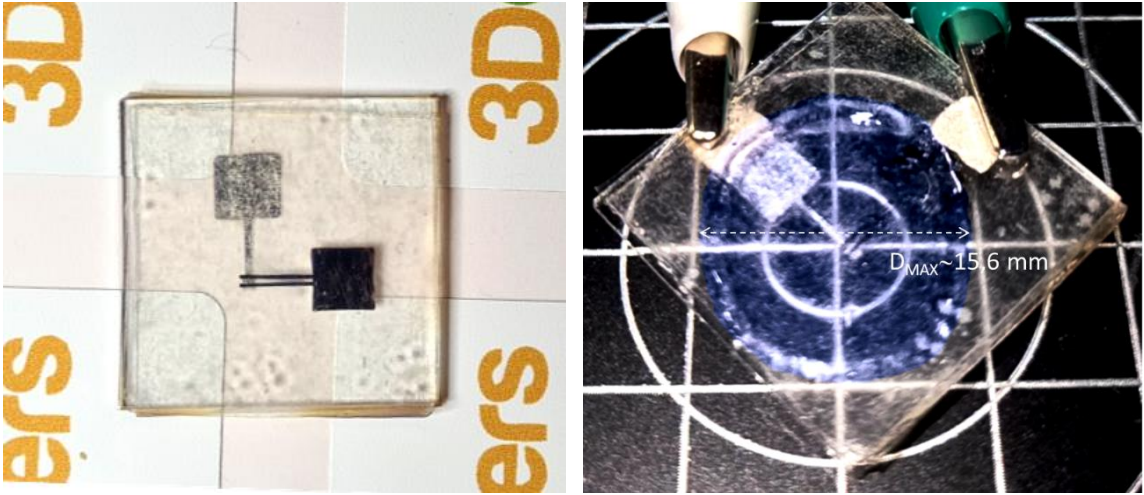

**Fig. S2 | Photograph of the device and maximum spreading diameter ($D_{MAX}$).** Left) Photograph of a single Trecxel device. Right) Colored snapshot recorded during spreading of the drop at the impact with the surface.

As shown in Equation 1 in the main text, the voltage generated by the Trecxel is proportional to the time derivative of the Area ($A_e(t)$). It needs to be taken into account that the active area of the Trecxel that is overlapped by the drop is a rectangle with a defined area of $A = L \cdot (L - 2W)$. Assuming a perfect impact of the drop on the center of the Trecxel, the spreading of the drop presents two regimes: i) the first corresponding to the area of the circle $A = \pi \cdot r(t)^2$, which occurs up to the moment t=5.3 μs in which $r(t) = \frac{L}{2} = 0.25 \, mm$ and ii) the second that corresponds to the filling of the rest of the rectangle.

In the regime i), we can express the relation shown in Equation EE1, as the area of a circle:

$$A_E(t) = \frac{\pi}{4} \frac{3 D_{MAX}^2 V}{8 D_0} t \; [m^2] \quad \text{(Equation S2)}$$

One remarkable characteristic of this relation shown in Equation S2 is that the time derivative of the area is:

$$\frac{dA_E(t)}{dt} = \frac{\pi}{4} \frac{3 D_{MAX}^2 V}{8 D_0} = 36.6 \cdot 10^{-3} \; \left[ m^2/s \right] \quad \text{(Equation S3)}$$

This would indicate that, at first approximation, the power delivered by the Trecxel is constant and independent of time during the first regime of the spreading.

During the second regime (ii), the overlapping area between the active area and the area of the drop follows the relation:

$$A = r(t)^2 \left[ \pi - 2 \cdot arccos\left(\frac{L-2W}{2 \cdot r(t)}\right) + sin\left(2 \cdot arccos\left(\frac{L-2W}{2 \cdot r(t)}\right)\right) \right] \ [m^2] \quad \text{(Equation S4)}$$

During this regime, the time derivative of the area decreases and is zero when the entire area of the Trecxel is covered (approximated here as the instant when the drop border reaches the longest side of the rectangle, r(t)=L/2, neglecting the corners). The time derivative of the area of the Trecxel covered by the drop during spreading is shown in the inset of the Fig. S3 (red curve).

As a result of this sudden increase at the water-solid interface, the output voltage rises rapidly, as shown in Fig. 1b in the main text and Fig. S3 (gray curve). It is important to note that if we assume an uncertainty of 1 mm at the impact, the spreading time of the drop within the active area of the Trecxel is increased in one order of magnitude to 120-160 µs.

Once the active area is covered by the droplet, the device does not deliver any voltage, and the spreading dynamic stops contributing to the actuation of this individual Trecxel. As presented in the main text, a great advantage of the Trecxel arrays would be that the remaining kinetic energy from such a droplet that spreads to diameters higher than 15 mm can be harvested by a neighboring Trecxel.

*Supplementary Information 1.2: Dynamics for a Recoiling Droplet.*

Immediately after impact, the drop slides (as the device is tilted by 30º and the surface is hydrophobic), and after approximately 2-3 ms, the drop undergoes the recoiling process, which gradually reduces the contact area of the water-solid interface. In the work of Guan *et al.,*[5] the authors reported similar experimental conditions using tilted hydrophobic surfaces with water contact angles of 120º (although using drops with lower speed and diameters) and studied both the bouncing and sliding of the drops.

According to their studies, at the moment of the impact, the drop starts to increase in size until reaching $D_{MAX}$ after approximately 2-3 ms. After this point, most of the authors report an almost linear recoiling rate that for similar experimental conditions can be approximated as:

$$\frac{D(t)}{D_0} = -10^3 \frac{t}{2} + 2.7 \quad \text{(Equation S5)}$$

(the constant term varied in their experiments from 2.4 to 3.0).[5]

And consequently, the radius from the center of the drop can be written as:

$$r(t) = \frac{-10^3}{4} D_0 \cdot t + \frac{2.7}{2} D_0 \quad \text{(Equation S6)}$$

At the same time, the center of the drop slides at approximately a constant speed. According to previous works,[5] the sliding speeds for a hydrophobic surface and a drop impacting with a hydrophobic surface at 2.4 m/s is around $v_{slide}$~0.5 m/s. We can assume that the front of the drop is approximately flat, which means that the radius of the drop is much bigger than the size of the Trecxel. Under these assumptions, the position of the front of the drop and the area of the Trecxel covered can be described by:

$$x_{front}(t) = r_{recoil}(t) + r_{slide}(t) = \left(\frac{-10^3}{4} D_0 \cdot t + \frac{2.7}{2} D_0\right) - (0.5 \cdot t) \quad \text{(Equation S7)}$$

$$A_E(t) = (L - 2W) \cdot \left[\left(\frac{-10^3}{4} D_0 - 0.5\right) t + \left(\frac{2.7}{2} D_0 + \frac{L}{2}\right)\right] m^2 \quad \text{(Equation S8)}$$

Taking into account this relation, the front of the drop would touch the border of the active area at 3.53 ms from the impact ($x_{front} = 0.45 \cdot 10^{-3} m$ if the Trecxel metallic fingers are aligned with the movement of the drop). From t=3.53 to 4.06 ms, the dynamic of the droplet would produce a decrease in the active area of the Trecxel, given by:

$$\frac{dA_E(t)}{dt} = (L - 2W) \cdot \left(\frac{-10^3}{4} D_0 - 0.5\right) m^2/s \quad \text{(Equation S9)}$$

As occurred for the regime i) during spreading, the time derivative of the active area covered by the droplet is constant from t=3.53 to 4.06 ms, as can be seen in Fig. S3 (red curve).

*Supplementary Information 1.3: Comparison with the voltage delivered by a single Trecxel.*

As discussed above, the kinetic energies of the droplet impact during the spreading and recoiling processes result in the modification of the active area and, consequently, affect the voltage and power generated by the device, presenting different time-scale characteristics for both processes. Fig. S3 shows a comparison of the voltage pulse generated by a single Trecxel (freshly charged, grey curve) and the $\frac{dA_E(t)}{dt}$ that overlaps with the active area of the Trecxel, as calculated above (red curve). It can be noted a very good agreement in the spreading dynamics regarding the time scale. Thus, the calculation based on the Pasandideh-Fard *et al.*[6] analytical model predicts a pulse duration of 17.2 µs, in a very good agreement with the observed voltage pulse. On the other side, the recoiling dynamics envisages a much less intense negative voltage pulse, that agrees with a barely visible negative voltage observed in the measurements.

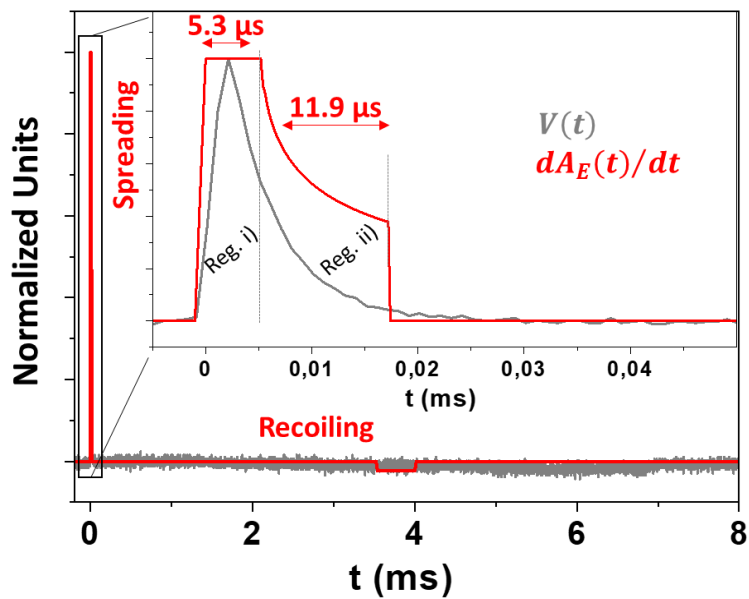

**Fig. S3| Comparison of the Voltage provided by a single Trecxel with the $dA_E(t)/dt$.** The experimentally acquired voltage for a PFA single-Trecxel pulse (grey curve) and the calculation of the time derivative of the active area overlapped with the drop (red curve). The inset shows a zoom in the positive voltage branch.

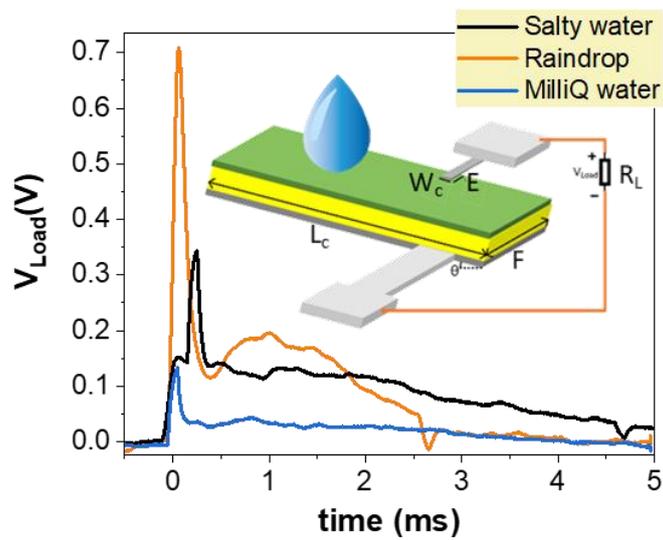

**Fig S4.** Output voltage of the control device for three different water drops and optimum loads. The inset shows the schematic of the control device with dimensions $L_c$=30 mm, $F$=20 mm, $W_c$=200 µm, and $E$=5 mm. The falling water drop first hits the dielectric layer, loses part of its kinetic energy, and then moves downward towards the active area. The orange, black, and blue lines correspond to salty, rain, and milli-Q water, respectively.

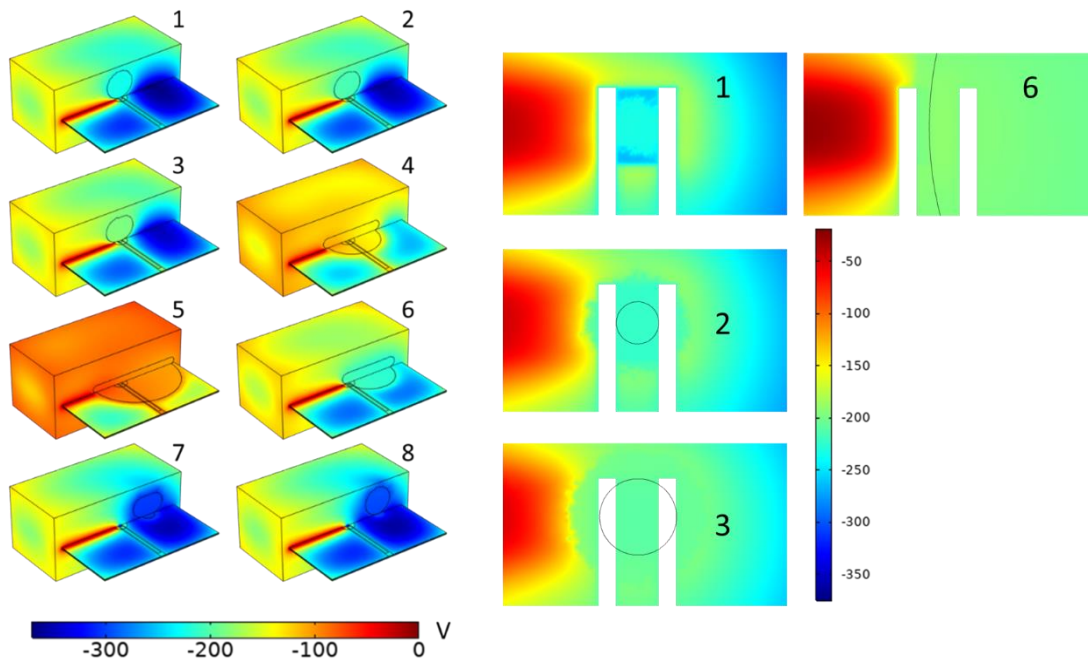

**Fig. S5| Snapshots of the simulated electrostatic potential distribution (in V) at different instants during the spreading and recoiling stages of the drop.** The PFA Surface has parasitic surface charges distributed uniformly

throughout the surface and in the middle of the top electrodes the density of negative charges is higher, as can be seen in the fig. 3(a) of the main text.

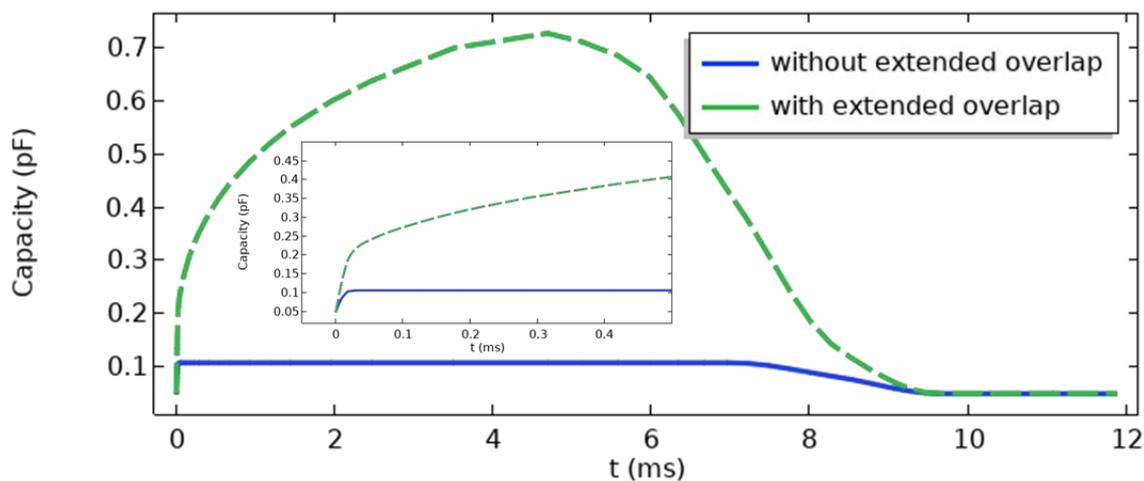

**Fig. S6| Time dependence of the Trecxel capacity.** The graph shows two cases: considering the overlap between the drop and the central part of the electrodes only and considering the entire overlap between the drop+top electrodes and all the bottom electrode. The inset shows a close up for the initial instants of the drop spreading.

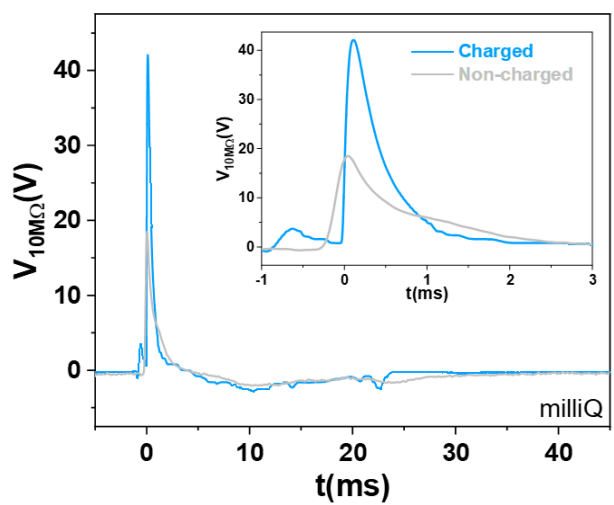

**Fig. S7| Comparison of the Voltage provided by a single Trecxel with and without e-beam activation of surface charges.** The experimentally acquired voltage for a PFA single-Trecxel for pristine (grey curve) and charged (blue curve) PFA after a similar droplet impact. The inset shows a zoom in of the positive voltage branch.

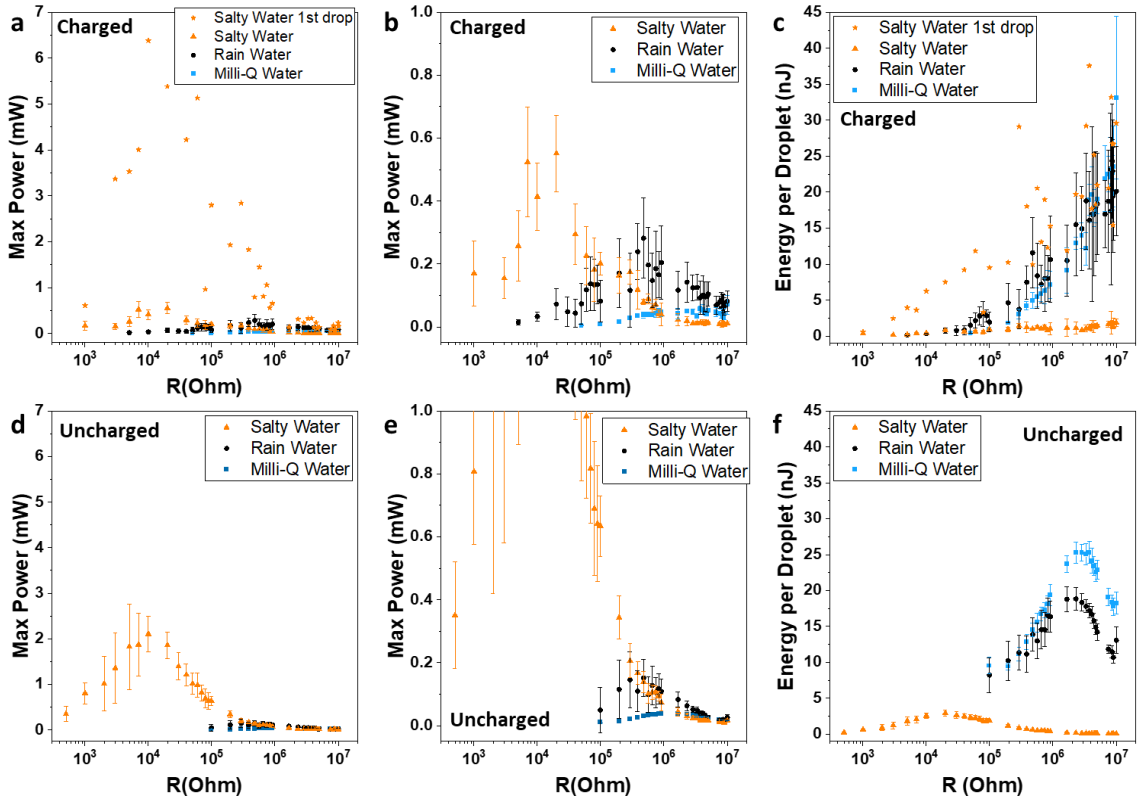

**Fig. S8| Power–load and Energy–Load curves for pre-charged and non-charged systems.** Maximum power (a, b, d, e) and Energy per droplet (c,f) versus load for charged (a-c) and non-charged (d-f) triboelectric layers for the three different types of droplets. Note that panels b and e are zooms of the panels a and d, respectively. The curves for charged devices includes the points corresponding to the first drop measurement with salty water (orange stars). Note that both, the Maximum power and Energy per droplet is higher using the first drop, indicating that the stored charged is decreased with every droplet impact.

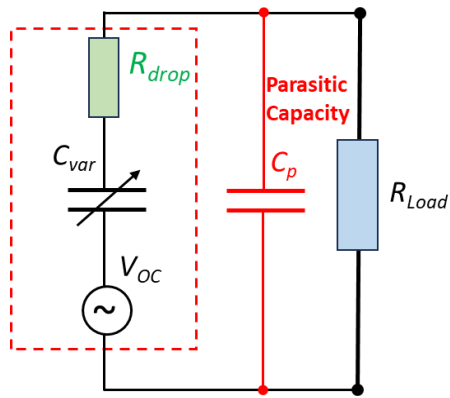

**Fig. S9| Equivalent TENG circuit of the simulated model including a parasitic capacitance.** The later is modeled as a capacitance connected in parallel to the original equivalent TENG and the load resistance.

## References

1. Lv, C., Hao, P., Zhang, X. & He, F. Drop impact upon superhydrophobic surfaces with regular and hierarchical roughness. Appl. Phys. Lett. 108, 141602 (2016).

2. Yun, S. Bouncing of an ellipsoidal drop on a superhydrophobic surface. Sci Rep 7, 17699 (2017).

3. Kumar, S., Namura, K., Suzuki, M. & P. Singh, J. Water droplet bouncing on a non-superhydrophobic Si nanospring array. Nanoscale Advances 3, 668–674 (2021).

4. Wang, H., Liu, C., Zhan, H. & Liu, Y. Droplet Asymmetric Bouncing on Inclined Superhydrophobic Surfaces. ACS Omega 4, 12238–12243 (2019).

5. Guan, Y., Fu, J., Wu, S., Chen, X. & Zhou, C. The post-impact dynamics of drop rebound on inclined hydrophobic surfaces of various wettabilities. Physics of Fluids 33, 072108 (2021).

6. M. Pasandideh-Fard, Y. M. Qiao, S. Chandre, and J. Mostaghimi, Phys. Capillary effects during droplet impact on a solid surface. Fluids 8, 650 (1996).

## Videos available from authors

**Video S1.** Single drop dynamics impacting a single Trecxel vs voltage output

**Video S2.** Switching on an LED by a single impact

**Video S3.** Activation of the Trecxel by continuous dropping.

**Video S4.** Drop impacting on a 2 x 2 Trecxel array.